\documentclass[
 reprint,
superscriptaddress,
preprintnumbers,
 amsmath,amssymb,
 aps,
floatfix,
]{revtex4-2}

\usepackage{ulem}
\usepackage{graphicx}
\graphicspath{ {./images/} }
\usepackage{dcolumn}
\usepackage{bm}
\usepackage{xcolor}
\usepackage{hyperref}
\usepackage[mathlines]{lineno}
\usepackage{siunitx}
\usepackage{comment}

\begin{document}

\preprint{preprint for arXiv}

\title{Neutron detection and application with a novel 3D\textcolor{black}{-}projection scintillator tracker in the future long-baseline neutrino oscillation experiments} 

\newcommand{\BNL}{Brookhaven National Laboratory, Upton, NY, USA}
\newcommand{\CAU}{Chung-Ang University, Seoul, South Korea}
\newcommand{\CERN}{CERN, European Organization for Nuclear Research}
\newcommand{\CEA}{IRFU, CEA Saclay, Gif-sur-Yvette, France}
\newcommand{\ETH}{ETH Zurich, Zurich, Switzerland}
\newcommand{\INR}{INR, Institute for Nuclear Research, Moscow, Russia}
\newcommand{\LIS}{University of Lisbon, Lisbon, Portugal}
\newcommand{\LSU}{Louisiana State University, Baton Rouge, LA, USA}
\newcommand{\UMD}{University of Minnesota, Duluth, MN, USA}
\newcommand{\Penn}{University Pennsylvania, Philadelphia, PA, USA}
\newcommand{\Pitt}{University of Pittsburgh, Pittsburgh, PA, USA}
\newcommand{\Rochester}{University of Rochester, Rochester, NY, USA}
\newcommand{\SDSMT}{South Dakota School of Mines and Technology, Rapid City, SD, USA}
\newcommand{\SB}{Stony Brook University, Stony Brook, NY, USA}
\newcommand{\WM}{College of William and Mary, Williamsburg, VA, USA}
\newcommand{\MG}{University of Antananarivo, Madagascar}
\newcommand{\UCB}{University of California, Berkeley, CA, USA}
\newcommand{\LBL}{Lawrence Berkeley National Laboratory, CA, USA}

\newcommand{\MIPT}{Moscow Institute of Physics and Technology (MIPT), Moscow, Russia}
\newcommand{\MEPhl}{Moscow Institute of Engineering and Physics (MEPhl), Moscow, Russia}

\newcommand{\LPI}{LPI, Lebedev Physics Institute, Moscow, Russia}

\author{S.~Gwon}            \affiliation{\CAU}
\author{P.~Granger}                     \affiliation{\CEA}
\author{G.~Yang}\thanks{Corresponding author, gyang9@berkeley.edu}                        \affiliation{\UCB}\affiliation{\LBL}
\author{S.~Bolognesi}                   \affiliation{\CEA}
\author{T.~Cai}                         \affiliation{\Rochester}
\author{M.Danilov}          \affiliation{\LPI}
\author{A.Delbart}          \affiliation{\CEA}
\author{A.~De~Roeck}            \affiliation{\CERN}
\author{S.~Dolan}           \affiliation{\CERN}
\author{G.~Eurin}           \affiliation{\CEA}
\author{R.F.~Razakamiandra}      \affiliation{\MG}
\author{S.~Fedotov}                     \affiliation{\INR}
\author{G.~Fiorentini~Aguirre}      \affiliation{\SDSMT}
\author{R.~Flight}          \affiliation{\Rochester}
\author{R.~Gran}            \affiliation{\UMD}
\author{C.~Ha}                          \affiliation{\CAU}
\author{C.K.~Jung}                      \affiliation{\SB}
\author{K.Y.~Jung}          \affiliation{\CAU}
\author{S.~Kettell}         \affiliation{\BNL}
\author{M.Khabibullin}  \affiliation{\INR} 
\author{A.~Khotjantsev}         \affiliation{\INR}
\author{M.~Kordosky}            \affiliation{\WM}
\author{Y.~Kudenko}         \affiliation{\INR} \affiliation{\MIPT} \affiliation{\MEPhl}
\author{T.~Kutter}          \affiliation{\LSU}
\author{J.~Maneira}         \affiliation{\LIS}
\author{S.~Manly}                       \affiliation{\Rochester}
\author{D.A.~Martinez~Caicedo}        \affiliation{\SDSMT}
\author{C.~Mauger}                      \affiliation{\Penn}
\author{K.~McFarland}           \affiliation{\Rochester}
\author{C.~McGrew}                      \affiliation{\SB}
\author{A.~Mefodev}         \affiliation{\INR}
\author{O.~Mineev}          \affiliation{\INR}
\author{D.~Naples}          \affiliation{\Pitt}
\author{A.~Olivier}                     \affiliation{\Rochester}
\author{V.~Paolone}                     \affiliation{\Pitt}
\author{S.~Prasad}          \affiliation{\LSU}
\author{C.~Riccio}                      \affiliation{\SB}
\author{J.~Rodriguez~Rondon}                   \affiliation{\SDSMT}
\author{D.~Sgalaberna}                  \affiliation{\ETH}
\author{A.~Sitraka}                 \affiliation{\SDSMT}
\author{K.~Siyeon}          \affiliation{\CAU}
\author{N.Skrobova} \affiliation{\LPI}
\author{H.~Su}                          \affiliation{\Pitt}
\author{S.Suvorov}   \affiliation{\INR}
\author{A.~Teklu}                       \affiliation{\SB}
\author{M.~Tzanov}          \affiliation{\LSU}
\author{E.~Valencia}            \affiliation{\WM}
\author{K.~Wood}                        \affiliation{\SB}
\author{E.~Worcester}           \affiliation{\BNL}
\author{N.Yershov}  \affiliation{\INR}

\date{\today}

\begin{abstract}
Neutrino oscillation experiments require a precise measurement of the neutrino energy. However, the kinematic detection of the final-state neutron in the neutrino interaction is missing in current neutrino oscillation experiments. The missing neutron kinematic detection results in a feed-down of the detected neutrino energy compared to the true neutrino energy. A novel 3D\textcolor{black}{-}projection scintillator tracker, which consists of roughly ten million 
active cubes covered with an optical reflector,
is capable of measuring the neutron kinetic energy and direction on an event-by-event basis using the time-of-flight technique thanks to the fast timing, fine granularity, and high light yield. 
The $\bar{\nu}_{\mu}$ interactions tend to produce neutrons in the final state. 
By inferring the neutron kinetic energy, the $\bar{\nu}_{\mu}$ energy can be reconstructed better, allowing a tighter incoming neutrino flux constraint.
This paper shows the detector's ability to reconstruct neutron kinetic energy and the $\bar{\nu}_{\mu}$ flux constraint achieved by selecting the charged-current interactions without mesons or protons in the final state.
\end{abstract}

\maketitle

\section{Introduction}
The future long-baseline neutrino oscillation experiments will accomplish a measurement of the CP-violating phase with unprecedented precision by measuring the difference in 
\textcolor{black}{oscillations to electron neutrinos and electron antineutrinos}~\cite{ND_CDR}.
The neutrino energies of the future long-baseline neutrino experiments range from hundreds of MeV up to a few GeV.
The neutrino interaction modes in this range are mainly charged-current quasi-elastic (CCQE) scattering, CC resonant scattering (RES), and CC deep-inelastic scattering (DIS). 
The neutrino cross section for those scattering modes has different energy dependence~\cite{Formaggio:2012cpf}. 
In order to discern the oscillation phenomena, the experiments reconstruct the neutrino energy in the detector via the CC interaction resultant visible particles. 

A near detector is needed to measure unoscillated neutrino spectra and constrain the systematic uncertainties such as neutrino flux, interaction cross section, and detector acceptance.
A stringent constraint on the flux and neutrino interaction cross section from the near detector is required to achieve a precise oscillation measurement. While it is relatively straightforward to reconstruct charged particles, neutrons present a particular challenge in neutrino event reconstruction.
Considering the neutrons share some significant portion of initial neutrino energy, it is very beneficial to detect neutron kinematics directly in particle detectors.

The 3D\textcolor{black}{-}projection scintillator tracker (3DST) is proposed to be \textcolor{black}{a} powerful near detector in future long-baseline experiments \cite{Blondel:2017orl, T2K:2019bbb, ND_CDR, Berns:2022fmq, Boyarintsev:2021uyw}. It is capable of detecting neutron kinematics on an event-by-event basis. 
In this manuscript, a flux constraint study is performed \textcolor{black}{that includes the neutron kinematics} as an application to demonstrate the potential of 3DST to constrain the flux uncertainty in long-baseline neutrino oscillation experiments.
The DUNE flux~\cite{DUNEflux} is taken as an example in this work due to its wide energy coverage.

The structure of the paper is as follows.
\textcolor{black}{Section~\ref{section2}} presents the key features of the 3DST detector.
\textcolor{black}{Section~\ref{section3}} describes the neutron kinematic measurement. 
\textcolor{black}{Section~\ref{section4}} shows a detailed description of the detector simulation setup.
\textcolor{black}{Section~\ref{sec:neutron detection performance}} details the neutron detection performance, including neutron and neutrino energy reconstructions and low-transverse-momentum ($\delta p_T$) event selection.
\textcolor{black}{Section~\ref{section6}} illustrates a neutrino flux constraint study.

\section{The 3D-projection scintillator tracker} 
\label{section2}
The role of the near detector in the long-baseline neutrino experiments is to constrain the neutrino flux and cross-section systematic uncertainties \textcolor{black}{and central values} that are applied to the far detector. A stringent systematic constraint requires an accurate measurement of the neutrino interaction at the near detector. 
DUNE can operate in forward horn current (FHC) and reverse horn current (RHC) modes, which mainly produce neutrinos and antineutrinos, respectively. This study focuses on the RHC mode.
Among the final-state particles, especially in the CCQE channel, the neutron is the most difficult one to reconstruct. In most cases, the missing neutron energy leads to a noticeably lower reconstructed neutrino energy than the true neutrino energy. 
Fig.~\ref{fig:neutronE_portion} shows the ratios of the averaged primary neutron energy to neutrino (top plot) and antineutrino (bottom plot) energy in different CC interaction modes from the GENIE generator~\cite{genie}. The average energy fractions carried by neutrons are about 3\% and 10\% in neutrino and antineutrino QE modes with energy below \SI{1}{GeV}, respectively. 
It is worth noting that for 10\% of the RES channel in the antineutrino mode, the energy fraction carried by neutrons reaches \textcolor{black}{above} 40\% at low energy.

\begin{figure}
    \includegraphics[scale=0.1]{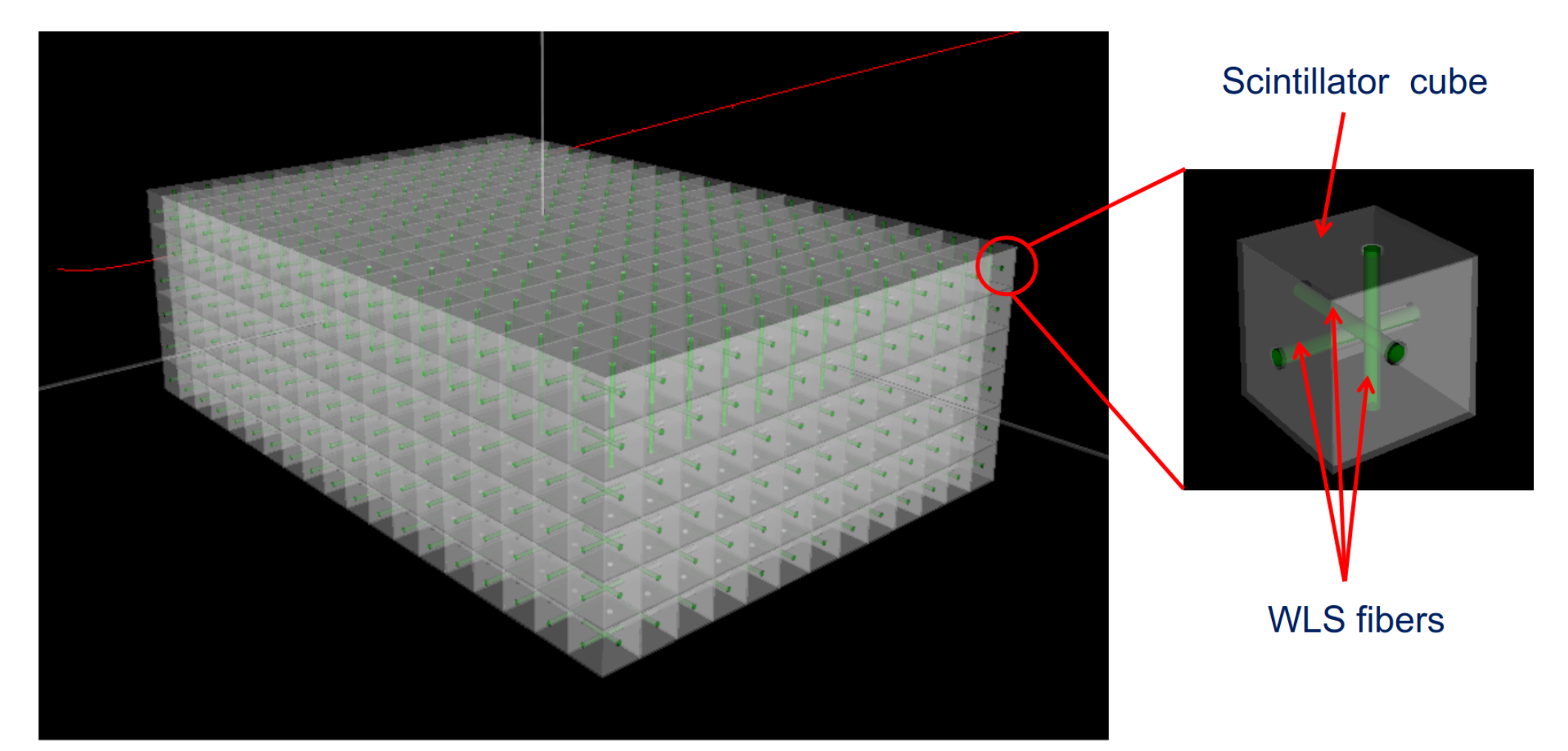}
    \caption{The concept of the 3D\textcolor{black}{-}projection scintillator tracker. Figure is taken from~\cite{T2K:2019bbb}.}
\label{3dst_concept}
\end{figure}
 \begin{figure}
    \includegraphics[scale=0.45]{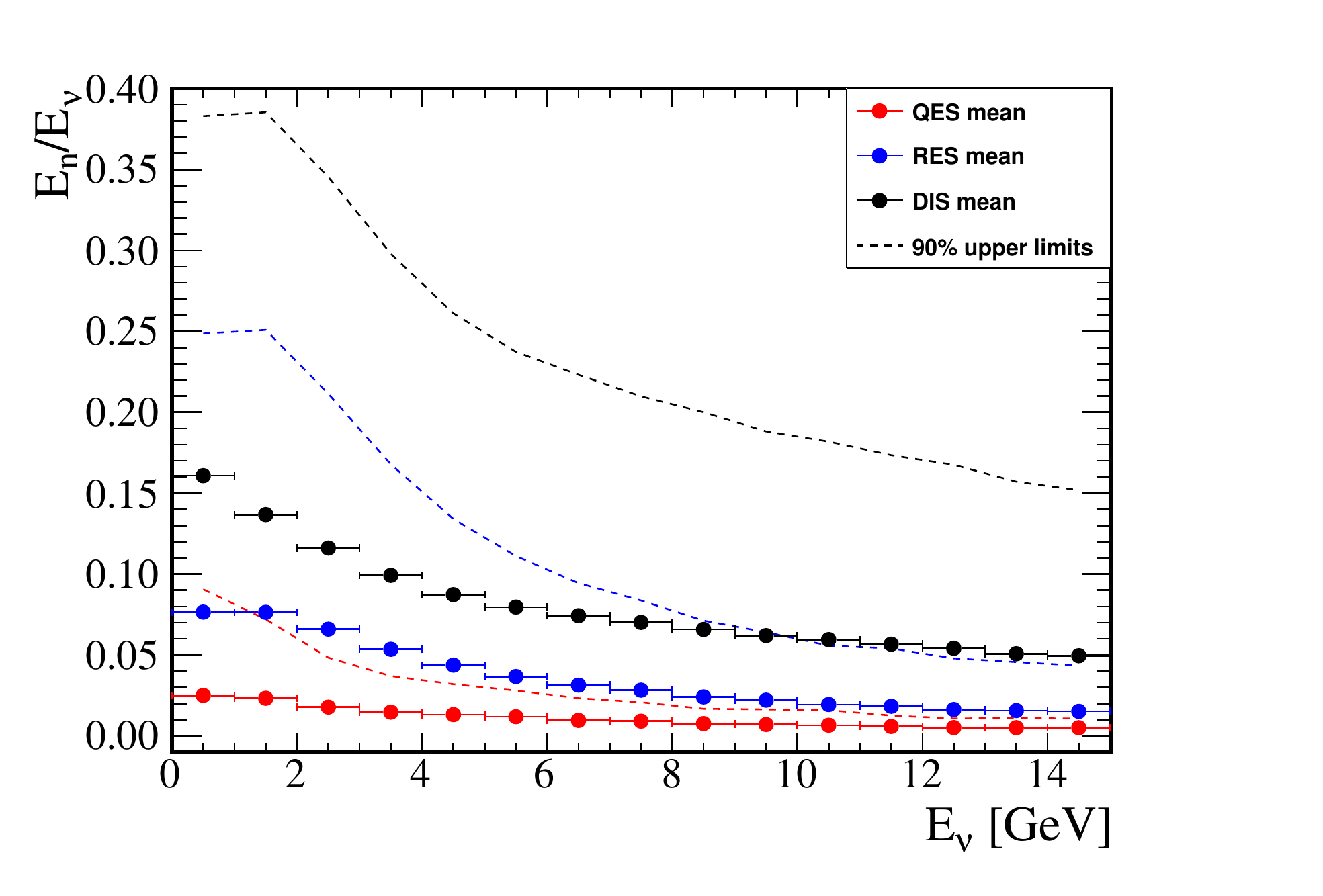}
    \includegraphics[scale=0.45]{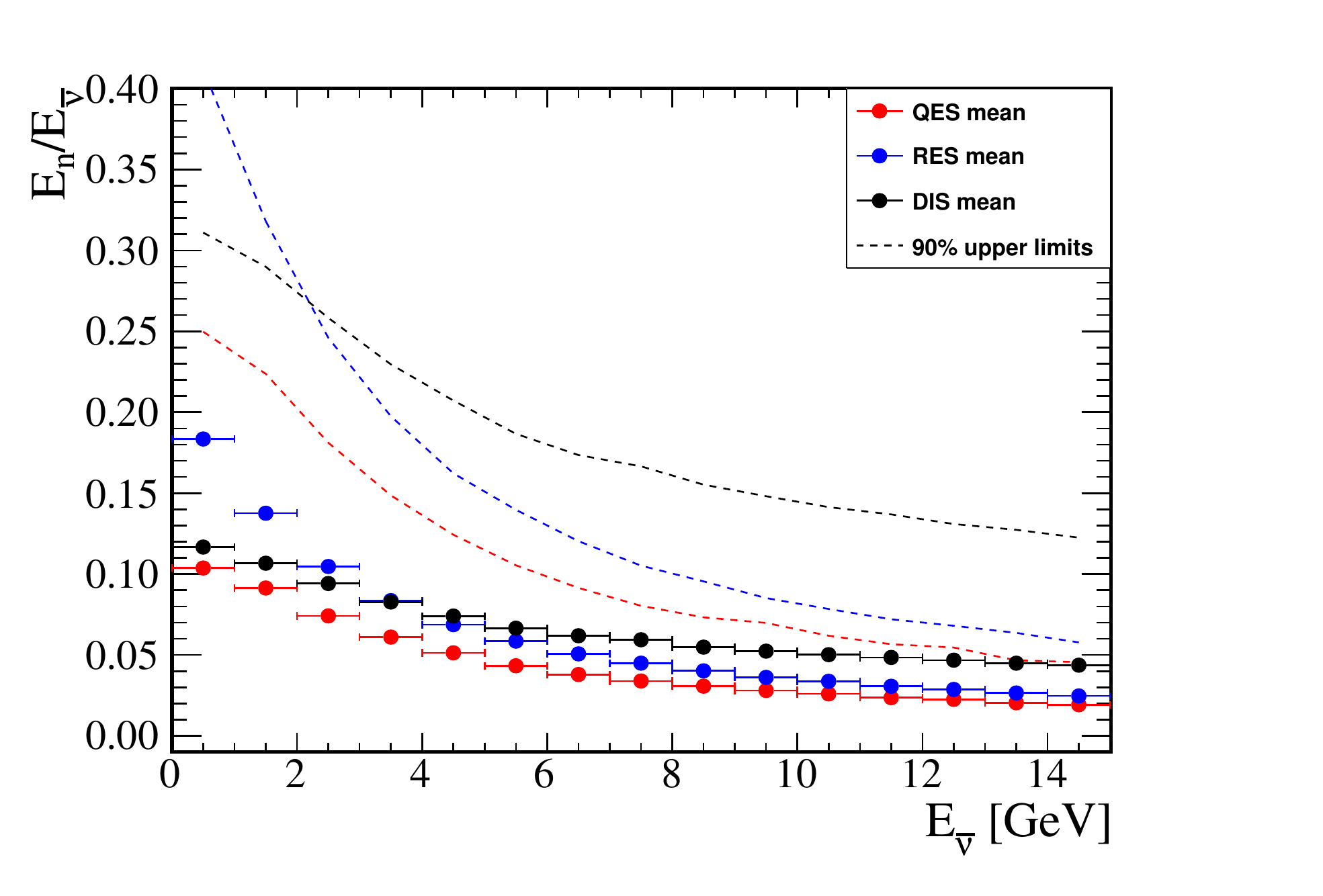}
    \caption{Average energy fraction transferred to the primary neutrons relative to the neutrino energy (top) and the antineutrino energy (bottom). The average ratios are calculated for CCQE, CCRES, CCCOH, and CCDIS interaction modes. 
    The dashed lines show for each channel the energy fractions below $90\%$ of the distribution.}
\label{fig:neutronE_portion}
\end{figure}
The original conceptual proposal of the 3DST detector can be found in~\cite{Blondel:2017orl}. The conceptual design of the detector is shown in Fig.~\ref{3dst_concept}. The detector consists of roughly ten million optically isolated plastic scintillator (CH) cubes with a total dimension of 2.4 m $\times$ 2.16 m $\times$ 1.92 m. The scintillation light inside each cube is absorbed by three wavelength-shifting fibers perpendicular to each other passing through the cube and read out by a SiPM at the end of each fiber. 
Three 2D-readout images of an event are constructed and combined to form a pseudo-3D image. The 3DST is characterized by the following features:
\begin{itemize}
\item Fine granularity with 1.5 $\times$ 1.5 $\times$ 1.5 cm$^3$ cube size and \textcolor{black}{a} fully active target;
\item 4$\pi$ solid angle acceptance giving momentum reconstruction even for the low momentum tracks;
\item Fast timing, 0.9 ns for each fiber and 0.5 ns for each cube (combining three fibers), suitable for detector neutrons~\cite{time_resolution}.
\end{itemize}
 
Due to the homogeneous and efficient performance of this detector, T2K has first adopted such a detector (SuperFGD) in its upgrade program. The SuperFGD is being fully built and will be a key component of the upgraded off-axis near detector ND280. In order to better characterize the detector, a SuperFGD prototype detector has been built. The prototype detector has a dimension of 24 cm $\times$  8 cm $\times$  48 cm with \textcolor{black}{a} 1 cm $\times$  1 cm $\times$  1 cm cube size. 
The prototype has been exposed to a charged particle beamline at CERN, and a significant amount of knowledge about the detector response was learned~\cite{T2K_charged_beamtest}. In addition, in light of the importance of neutron kinematic detection, two neutron beam tests were completed at the Los Alamos National Laboratory (LANL) in December 2019 and 2020. A large amount of neutron interaction data with energy ranging from 0 to 800 MeV has been collected with the prototype detector. With the LANL beam test data, the detector response to neutrons can be understood in detail. 
It demonstrated the individual neutron kinematics detection capability of the prototype by measuring the n-CH total cross section.
The main results have been published in~\cite{neutron_paper}.
In addition, the beam test program is providing the neutron detection efficiency, scattering angle, secondary scattering rate, and exclusive channel production information\textcolor{black}{, such as pion or proton production,} as functions of the neutron energy.

Furthermore, the 3DST neutron kinematic detection application has been explored in previous works. 
One of the possible measurements is the transverse kinematic balance of final-state particles in the process of CCQE $\bar{\nu}_l ~p \rightarrow l^+ n$. When an antineutrino interacts with a target proton in the detector, if the proton is not bound, as in hydrogen, the sum of momenta in the plane perpendicular to the direction of incoming antineutrino, denoted by $\delta p_T$, vanishes. Non-zero $\delta p_T$ implies that the final-state particles are not free from Fermi motion, binding energy, or \textcolor{black}{final-state interactions} (FSI) inside the nucleus~\cite{dpt_paper}. Thus, the $\delta p_T$ provides a powerful enhancement for the $\nu$-p sample selection. A comprehensive study of the transverse kinematics balance in the context of the SuperFGD detector is presented in~\cite{dpt_paper}. 
In addition, more recently, the impact of the SuperFGD neutron detection capability on the neutrino interaction cross section and flux constraint has been studied quantitatively~\cite{https://doi.org/10.48550/arxiv.2203.11821}. 
Another possible physics application with 3DST is neutrino flux constraint with the CC0$\pi$0p1n channel, which has a $\mu+$ and a neutron in the final state.
The CC0$\pi$0p1n channel has a relatively small model uncertainty compared to other channels due to the simple event topology.
The channel is selected in this work in order to show the possibility of constraining the incoming neutrino flux uncertainty with neutron kinematic detection\textcolor{black}{, as illustrated in Section~\ref{section6}}.

\textcolor{black}{It is worth noting that the CCQE and the CC0$\pi$0p1n channel are not identical. The CCQE channel is a type of true interaction channel, while the CC0$\pi$0p1n is a topological channel at the analysis level. We select the CC0$\pi$0p1n channel mainly intending for the CCQE interaction. Non-CCQE events can also contribute to the selected CC0$\pi$0p1n sample.}
 
\section{Neutron energy measurement}
\label{section3}
This work mainly explores the impact of the 3DST detector in the DUNE near detector hall with a distance of 574 m from the proton beam target. 
For simplicity, we focus on the CC0$\pi$ interaction in this paper. Topologically, the CC$0\pi$ channel is dominated by the CCQE and 2p2h events, and the main target channel is CCQE.
The CCRES interactions also contribute to CC0$\pi$ if the pions are absorbed in the nucleus.
\textcolor{black}{The relative fractions of CCQE, 2p2h, and others are 22.8\%, 8.5\%, and 68.7\%, respectively.}

Additionally, we try to select events with no protons and only one neutron in the final state.

The work presented in this paper strongly depends on the neutrino-interaction, nuclear and particle-propagation modeling. A few caveats should be clearly stated here.
\begin{itemize}
    \item The nuclear modeling uncertainties in the CC$0\pi$0p1n channel are covered in an approximated way by present studies. This study is an example demonstration of the 3DST capability.
    \item Both out-of-fiducial external background and internal background\textcolor{black}{s} are critical. The background estimate depends on the neutrino interaction modeling on the fiducial material and out-of-fiducial material. The detail of handling these backgrounds and the robustness of the background modeling, in particular in the CC0$\pi$0p1n channel study, is discussed in later chapters. 
    \item Systematic uncertainty due to FSI and secondary interaction (SI) may not be fully covered by this study. This problem is still under study by the community~\cite{FSIstudy}, and no robust uncertainties are available yet.
\end{itemize}

In neutrino experiments, neutrino energy reconstruction is performed mainly in three methods with the presence of neutrons:
\begin{enumerate}
    \item Using kinematics of the lepton: the energy and scattering angle of the lepton in the CCQE interaction is used, assuming momentum conservation for the two-body interaction. However, other channels can mimic the CCQE signature, thus causing significant bias.
    \item Summing up all energy deposits inside the fiducial volume: neutrons carry more energy in the form of kinetic energy than the energy deposit. The ``feed-down'' of the reconstructed energy can be significant.
    \item Measuring kinetic energy of the neutron: the neutron kinetic energy is estimated by measuring the neutron-induced isolated hit's time and distance to the neutrino interaction vertex. In the CCQE interaction, the neutron kinetic energy is added to the $\mu^+$ energy.
\end{enumerate}
The last method is \textcolor{black}{the} so-called Time-of-Flight method (ToF).

\begin{figure}[h]
\centering
\includegraphics[scale=0.2]{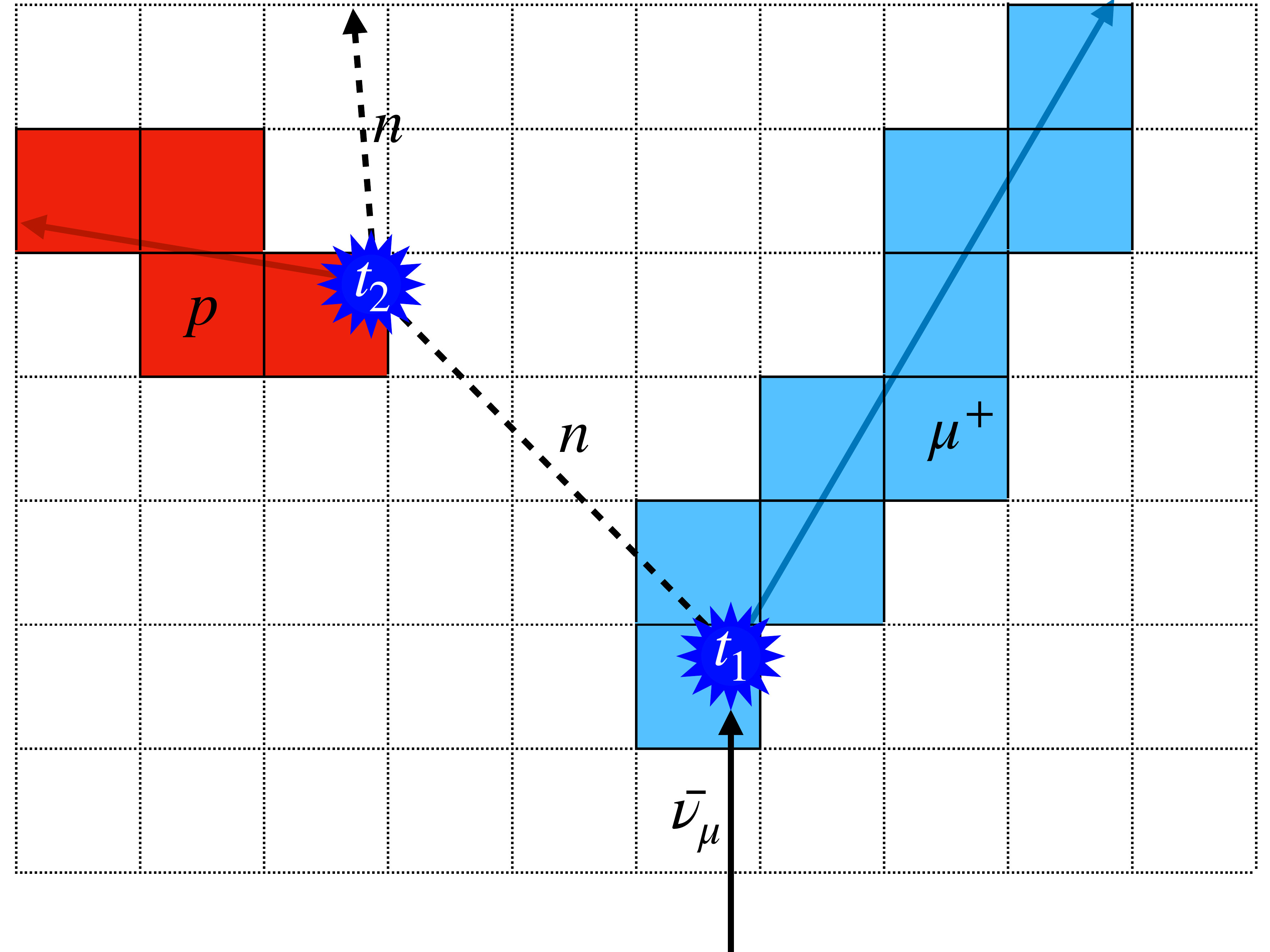}
\caption{Time-of-flight and lever arm. A neutron-induced cluster in \textcolor{black}{the} 3DST is identified by the first cluster after the $\bar{\nu}_\mu$ interaction.}
\label{fig:neutron_detection}
\end{figure}
The ToF technique for the neutron kinetic energy estimation is illustrated in Fig.~\ref{fig:neutron_detection}. For \textcolor{black}{a} $\bar{\nu}_{\mu}$ CCQE event in the 3DST, the start of a $\mu^+$ track at time $t_1$ is marked as the $\bar{\nu}_{\mu}$ interaction point. Then a cluster of signals occurring at a certain distance from the $\bar{\nu}_{\mu}$ interaction vertex at time $t_2$ is marked as the neutron interaction point. In Fig.~\ref{fig:neutron_detection}, the cluster in red represents a proton recoil. The time difference $t_2-t_1$ is the neutron time-of-flight, and the distance between the two interaction points is called the lever arm. For the CC$0\pi$ channel, we expect primary neutrons to be the main source of the isolated clusters. Selecting the first cluster in time allows us to pick up the primary neutron's first interaction, thus measuring its energy with the travel time and distance. 

\section{simulation setup}
\label{section4}
All analysis in this paper is based on a fully reconstructed Monte Carlo (MC) simulation. 
The expected DUNE flux in antineutrino beam mode is used in the MC simulation.
The GENIE generator v3.00.04 tune G1810a~\cite{genie} is used to model the neutrino interaction with the nucleus in the detector.
The modeling of the final-state particle propagation in the detector was completed by the \textit{edep-sim} package, which is a wrapper of the GEANT4 software~\cite{GEANT4}.
A realistic detector geometry is generated by DUNENDGGD package \cite{DUNE_ndggd}.
The full size of the detector is 2.4 m $\times$ 2.4 m $\times$ 1.95 m.
The simulation for the signal response of the detector, including the signal readout, DAQ, and calibration, is completed by the \textit{erep-sim} package~\cite{erep-sim}.

As a final step of the simulation chain, a full event reconstruction for the detector is performed by the CubeRecon package~\cite{cuberecon}.
For each event, the particle trajectories are projected into three 2D views, which contain each fiber's energy and time readouts.
The three 2D views are converted into 3D reconstructed objects.
There are two object classes: track\textcolor{black}{s} and cluster\textcolor{black}{s}.
\textcolor{black}{A} track is an object longer than three voxels; otherwise, the object is a cluster.
The objects have all the hit information, such as the position, charge, and time.
The following analysis is performed with the fully reconstructed objects.

\section{Neutron detection performance}
\label{sec:neutron detection performance}
In this \textcolor{black}{section}, we investigate the neutron detection performance of 3DST as well as the impact of this performance on the neutrino energy measurement. 
This study focuses on neutron detection in the CC0$\pi$ $\bar{\nu}_\mu$ interactions.
In addition to the full reconstruction, the following assumptions are made:
\begin{itemize}
    \item The particle identification (PID) of charged particles is assumed to be perfect, given the excellent reported $dE/dx$ resolution of \textcolor{black}{the} 3DST~\cite{T2K_charged_beamtest}.
    \item A muon momentum resolution of 4\% is applied. This resolution is realistic given the typical momentum resolutions reachable by spectrometers that would be placed around \textcolor{black}{the} 3DST~\cite{T2KTPC}.
    \item An angular resolution of \SI{1}{\degree} is applied for the azimuthal and polar angles of the muon, given the granularity of the detector.
\end{itemize}

\textcolor{black}{In this section, a few timing resolutions are assumed and compared. The expected timing resolution for the single channel readout is 0.9 ns, dominated by the scintillation time~\cite{timingResolution}. For each cube, the signal is read out by three channels thus the timing resolution can go down to $\frac{0.9}{\sqrt{3}} \approx 0.5$ ns. Typically, the neutron-induced cluster results in more than one cubes hence the timing resolution can go further down to $\frac{0.5}{\sqrt{N}}$ ns, where $N$ is the number of fired cubes. The electronics for the detector should be chosen to have a smaller impact on the timing resolution.  
}

Following these considerations, for each simulated event, the analysis strategy is the following :
\begin{enumerate}
    \item
    A single $\mu^+$ track with no additional tracks is considered.
    \item The first isolated reconstructed object in time is selected, assuming it corresponds to the first interaction of the primary neutron inside the detector.
    \item Topological cuts are applied to remove events not associated with the interaction of a primary neutron.
    \item The neutron momentum is estimated with the measured lever arm and time-of-flight.
\end{enumerate}

\begin{figure}[h!]
\centering
\includegraphics[scale=0.3]{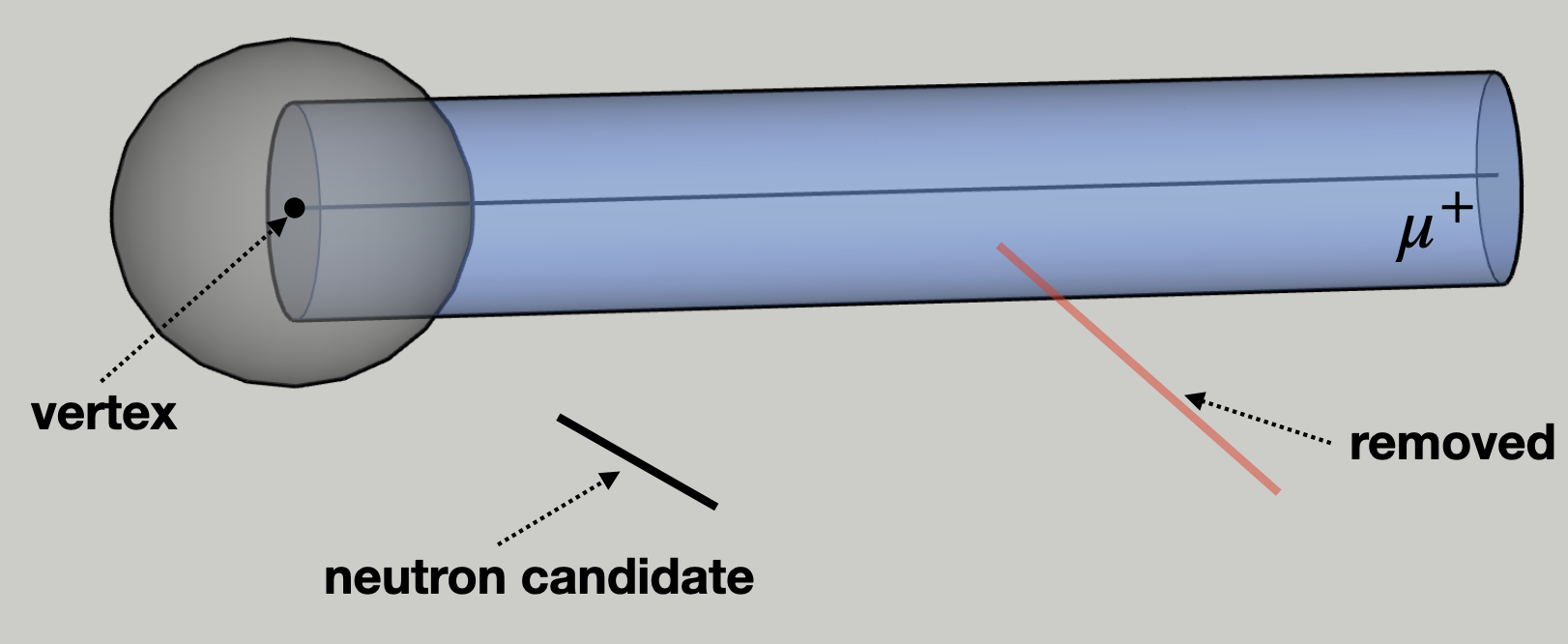}
\caption{An example of a single-track event. If the event has multiple tracks within the gray sphere, the event is rejected. The blue-colored track is the $\mu^+$ track. \textcolor{black}{The red-colored object is likely originated from the muon and rejected. Among the remaining objects, the first in time is selected as the neutron-induced candidate.}}
\label{fig:isolated_object}
\end{figure}

The muon track starting point is taken as the neutrino interaction vertex. If there is more than one track within a spherical region centered at the vertex, the event is rejected.
The region is shown as the gray sphere in Fig.~\ref{fig:isolated_object}.
The radius of the sphere cut is set to 5.25 cm, which is the smallest distance that the vertex can be isolated from other tracks.
CC charged\textcolor{black}{-}pion production events could be mostly rejected by this cut since most $\pi^\pm$ tracks in the final state are close to the neutrino interaction vertex.
The events remaining after the rejection are defined as ``single-track'' events.

\begin{figure}[ht]
    \centering
    \includegraphics[width=\linewidth]{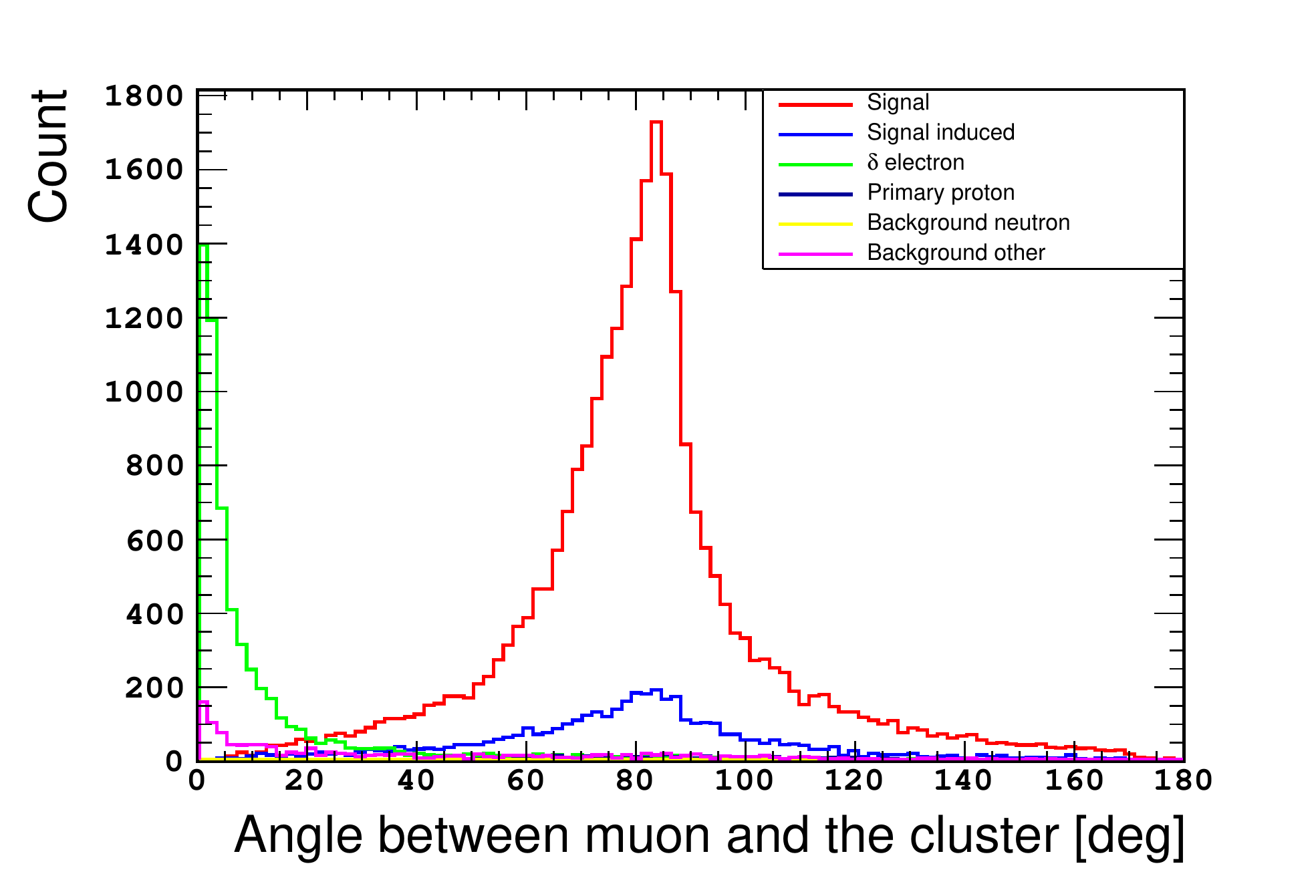}
    \caption{Distributions of the angular separation between the $\mu^{+}$ and the first cluster in time for the different kinds of particles.}
    \label{fig:cut_angle}
\end{figure}

For some events, the first isolated object in time is not induced by the primary neutron. In order to reject them, some additional selections are applied. For most of the cases where the first cluster in time is not related to the primary neutron, the energy deposit has been made by either a delta ray from the $\mu^{+}$ track, a secondary neutron created by the primary neutron or a primary proton produced by FSI. The secondary neutrons can hardly be distinguished from primary neutrons as they have similar topologies. It is, however, possible to remove most of the
delta ray electrons and primary protons by applying a cut on the angle between the $\mu^{+}$ track and the direction defined by the vertex and the first cluster, as shown in Fig.~\ref{fig:cut_angle}. 
Requiring an angle larger than \SI{30}{\degree} with the $\mu^{+}$ track allows us to increase the selection purity from 69\% to 81\% with a loss of only 2\% of signal.
Each category in Fig.~\ref{fig:cut_angle} is defined as follows;

\begin{itemize}
        \item Signal: Energy deposited by a primary neutron (by interacting with a proton, for instance);
        \item Signal induced: Energy deposited by a secondary neutron that acquired kinetic energy from an interaction with a primary neutron;
        \item $\delta$ electron: Energy deposited by a $\delta$ electron from the muon track;
        \item Primary proton: Energy deposited by a primary proton;
        \item Background neutron: Energy deposited by a neutron that was neither created in the primary interaction nor created by a primary neutron;
        \item Background other: Energy deposited by other kinds of particles, such as mesons.
 \end{itemize}

\begin{figure}
    \centering
    \includegraphics[width=\linewidth]{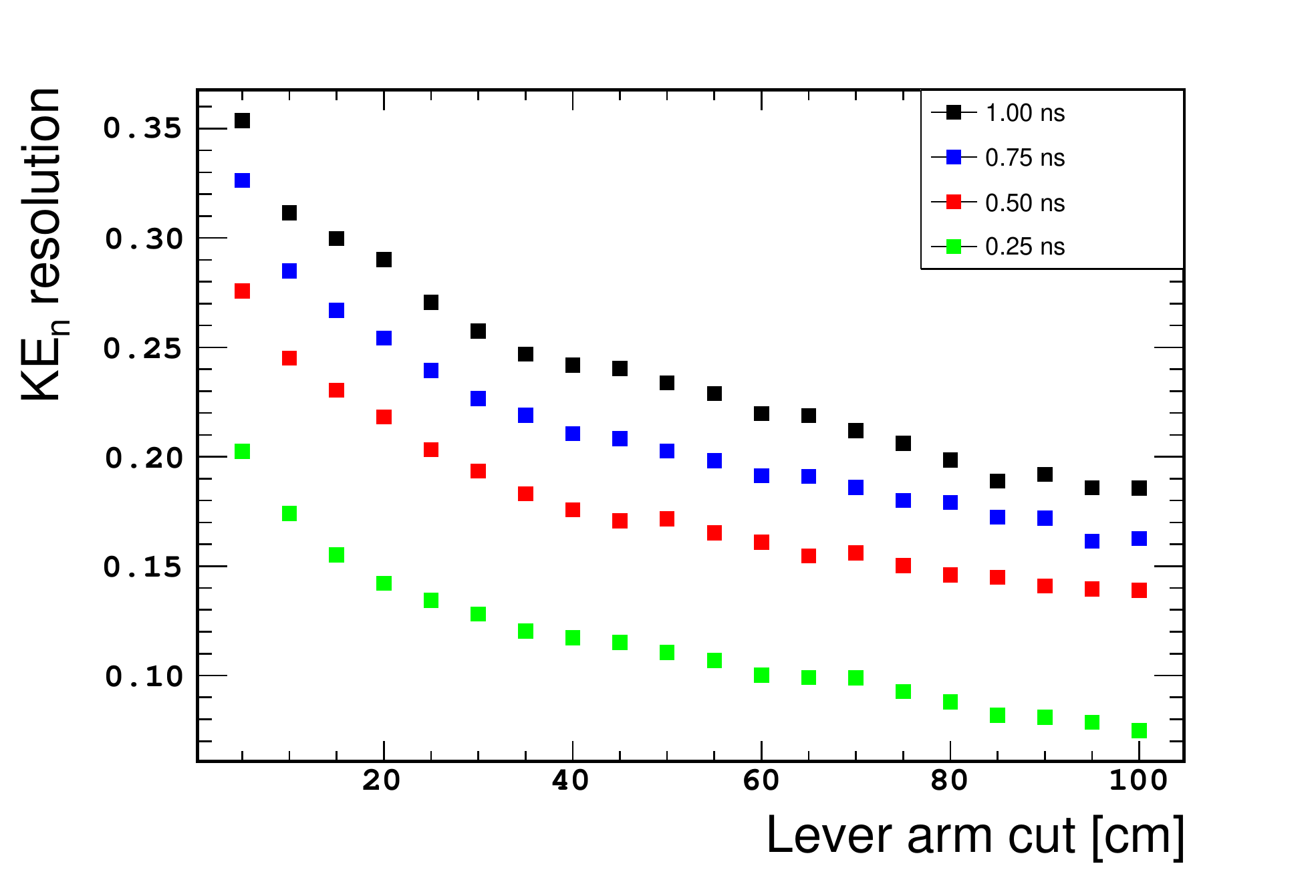}
    \caption{Neutron energy resolution as a function of the lever arm cut for various time resolutions.}
    \label{fig:neutronRes}
\end{figure}

Furthermore, an additional selection can be made by requiring a minimum distance between the antineutrino vertex and the earliest cluster (lever arm) in order to select a subset of events with neutrons that travel a sufficient distance. A longer lever arm results in a more precise energy reconstruction by ToF, given that the relative uncertainty on the lever arm decreases, leading to a better estimation of the neutron $\beta$, as reported in \cite{dpt_paper}. Fig.~\ref{fig:neutronRes} shows how the neutron kinetic energy resolution evolves as a function of the cut applied on the lever arm.  
Fig.~\ref{fig:neutronRes} demonstrates that improving the time resolution of such a detector allows for improving the neutron energy resolution.

Finally, as reported in \cite{dpt_paper}, an additional kinematic variable, the transverse momentum imbalance $\delta p_T$, allows us to select a subset of events for which the energy reconstruction is better controlled and as a consequence, the antineutrino energy resolution is improved.
In the case of a $\bar{\nu}$ CCQE interaction $\bar{\nu}p\rightarrow l^{+}n$, the transverse momentum imbalance is simply defined as:
\begin{equation}
    \delta p_T = \left|\vec{p^l}_T + \vec{p^n}_T \right|,
\end{equation}
where $p^n$ and $p^l$ are the outgoing neutron and lepton momenta, respectively. The $T$ subscript refers to the projection of the vector onto the plane transverse to the incoming neutrino direction.

\begin{figure}
    \centering
    \includegraphics[width=\linewidth]{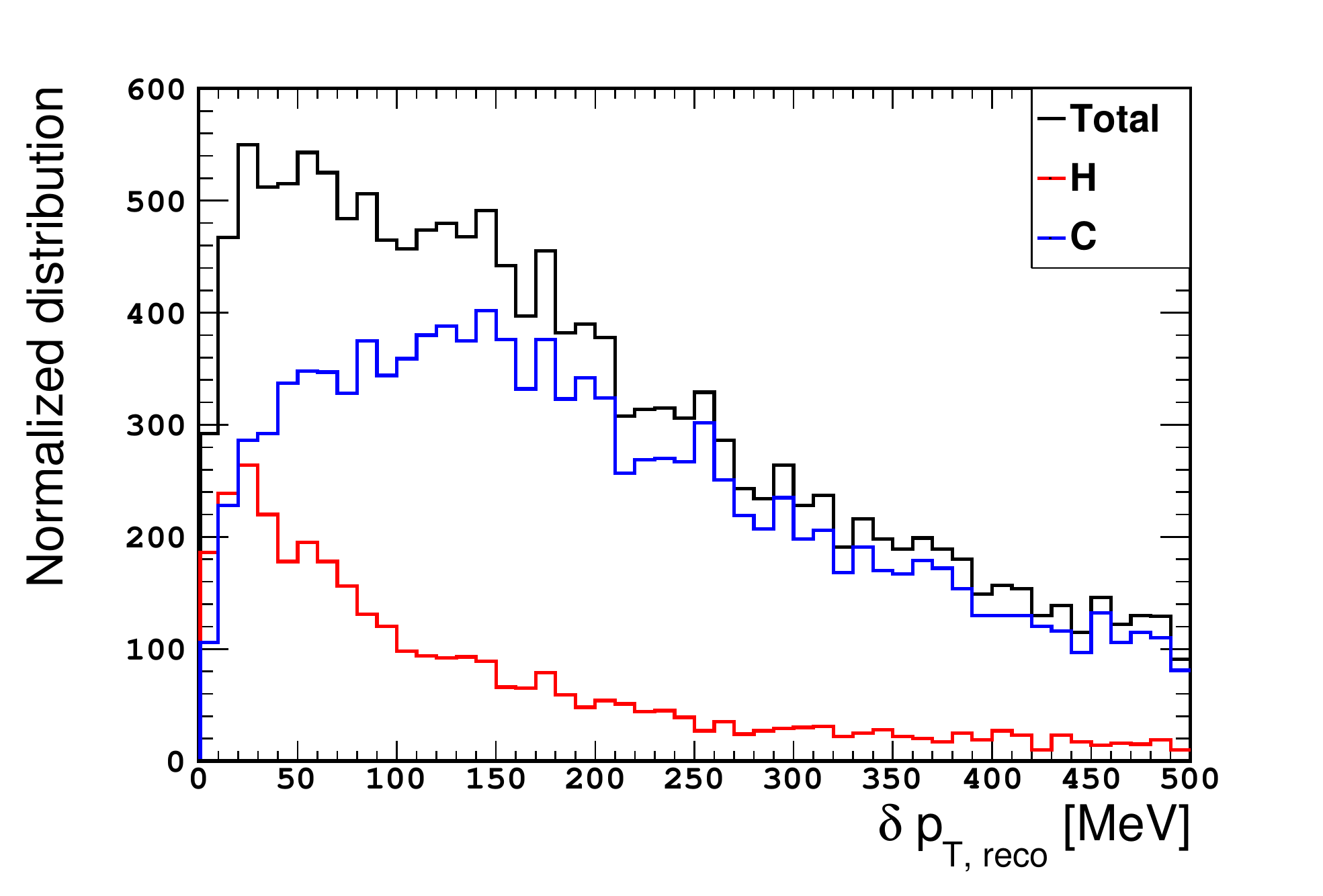}
    \caption{Reconstructed $\delta p_T$ distributions for interactions on Hydrogen and Carbon.}
    \label{fig:deltaPtReco}
\end{figure}

\begin{figure}
    \centering
    \includegraphics[width=\linewidth]{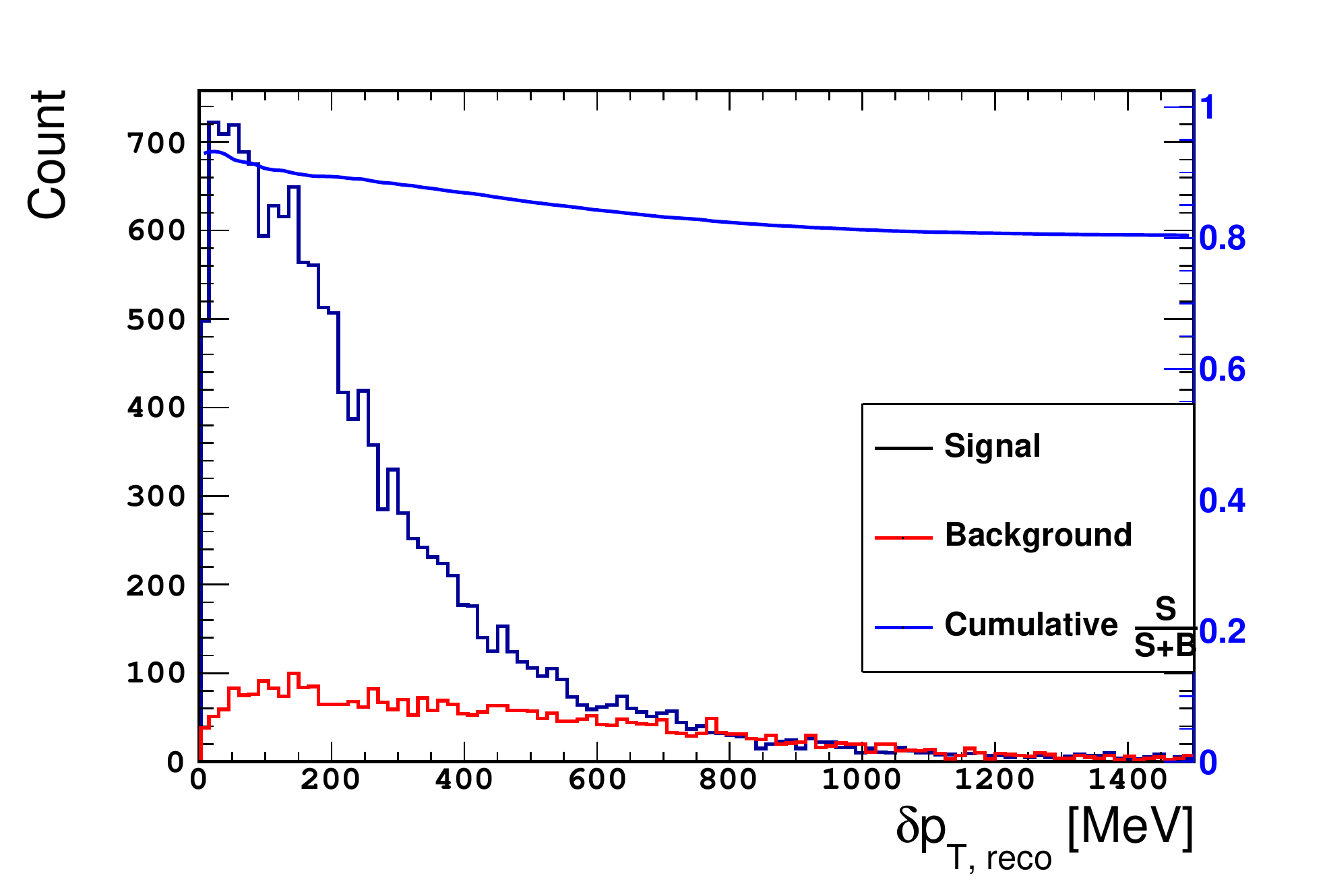}
    \caption{Signal and background distributions in the selected events as a function of the reconstructed $\delta p_T$. The blue curve gives the integrated signal fraction for events with $\delta p_{T, \text{reco}}$ below the considered value.}
    \label{fig:deltaPtBackground}
\end{figure}

There is no transverse momentum imbalance in the final state for an interaction on a free nucleon (the hydrogen target). On the other hand, nuclear targets subject to Fermi motion lead to a non-zero $\delta p_T$. Furthermore, inelastic neutrino interactions on nuclei with no meson in the final state can occur and \textcolor{black}{are difficult to distinguish} from elastic interactions. For example, 2p2h interactions or the production of a $\pi$ reabsorbed by the nucleus does not result in mesons in the final state. Consequently, selecting events with a low $\delta p_T$ allows the selection of a hydrogen-enriched sample. This can be seen in Fig.~\ref{fig:deltaPtReco} where the reconstructed $\delta p_T$ distributions for both hydrogen and carbon interactions are shown. 
Moreover, the interactions on carbon nuclei with a low $\delta p_T$ value tend to suffer less from the FSI and 2p2h. The unseen nucleon, in the case of 2p2h or absorbed pion, carries transverse momentum that is not measured and leads to the measurement of a large $\delta p_T$. In addition, applying a cut on $\delta p_T$ allows us to reject those events for which the primary neutron is misidentified or a meson is not reconstructed, as demonstrated in Fig.~\ref{fig:deltaPtBackground}. It can be seen that the application of a loose cut on $\delta p_T$, such as $\delta p_T < \SI{400}{MeV}$, can be enough to remove part of the background and enhance the selection purity from 81\% to 88\%
\textcolor{black}{ while a more stringent cut of \SI{40}{MeV} results in a purity of 93\%.
The efficiency for the neutrino hydrogen interaction with the $\delta p_T$ cut is from 10\% to 30\%, depending on the lever arm requirement~\cite{Munteanu_2020}.}

\begin{figure}
    \centering
    \includegraphics[width=\linewidth]{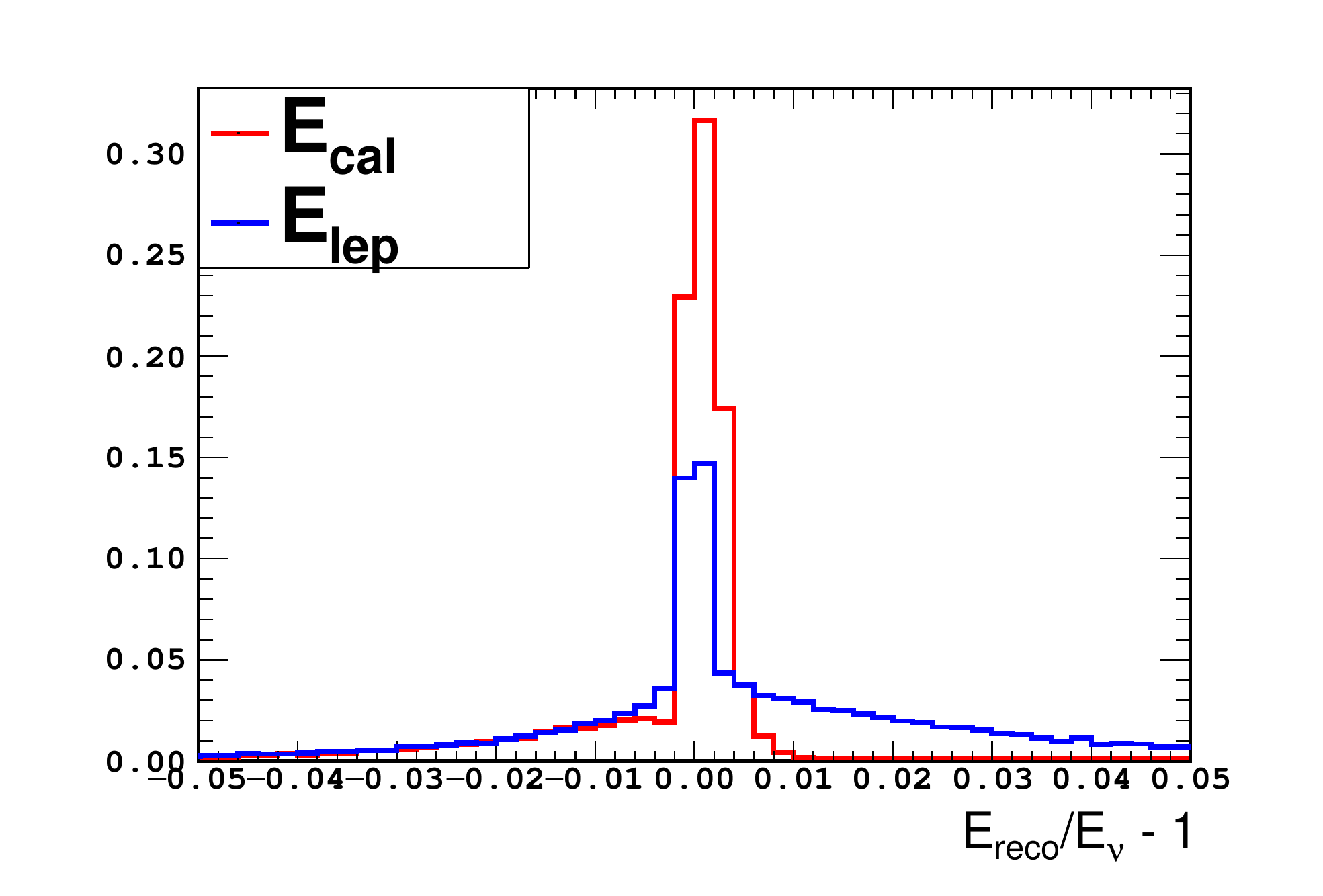}
    \caption{Expected neutrino resolutions for the two reconstruction methods assuming no detector smearing. The binding energy ($E_b$) is accounted for the reconstruction of $C$ interactions.}
    \label{fig:EnuResTRUE}
\end{figure}

For the CCQE antineutrino interactions, it is possible to rely only on the $\mu^{+}$ kinematics in order to compute the antineutrino energy:
\begin{equation}
    E_{\bar{\nu}}^{\text{lep}} = \frac{m_n^2 - m_p^2 - m_\mu^2 + 2m_p E_\mu}{2\left(m_p - E_\mu + p_\mu\cos{\theta_\mu}\right)}\label{eq:Elep},
\end{equation}
where $m_n$, $m_p$, and $m_\mu$ are the masses of the neutron, proton, and muon, respectively, whilst $E_\mu$, $p_\mu$, and $\theta_\mu$ are the energy, momentum, and angle of the outgoing $\mu^{+}$ with respect to the incoming antineutrino. This formula is accurate only in the case of an interaction on a free proton (hydrogen interaction). Fig.~\ref{fig:EnuResTRUE} shows the energy resolution with no detector smearing where there is a peak at zero corresponding to a perfect resolution for hydrogen interactions surrounded by a wide distribution due to the smearing caused by the Fermi motion.
Detecting the primary neutron of the antineutrino interaction allows us to better estimate the antineutrino energy by using a calorimetric measure of the total energy in the final state:
\begin{equation}
    E_{\bar{\nu}}^{\text{cal}} = E_\mu + E_n + (m_p - m_n)\label{eq:Ecal},
\end{equation}
where $E_\mu$ and $E_n$ are the muon energy and neutron energy measured with the spectrometer and ToF, respectively.
As shown in Fig.~\ref{fig:EnuResTRUE}, without detector smearing, the calorimetric energy reconstruction has a better resolution than the reconstruction without the neutron kinetic energy measurement.

\begin{figure}
    \centering
    \includegraphics[width=\linewidth]{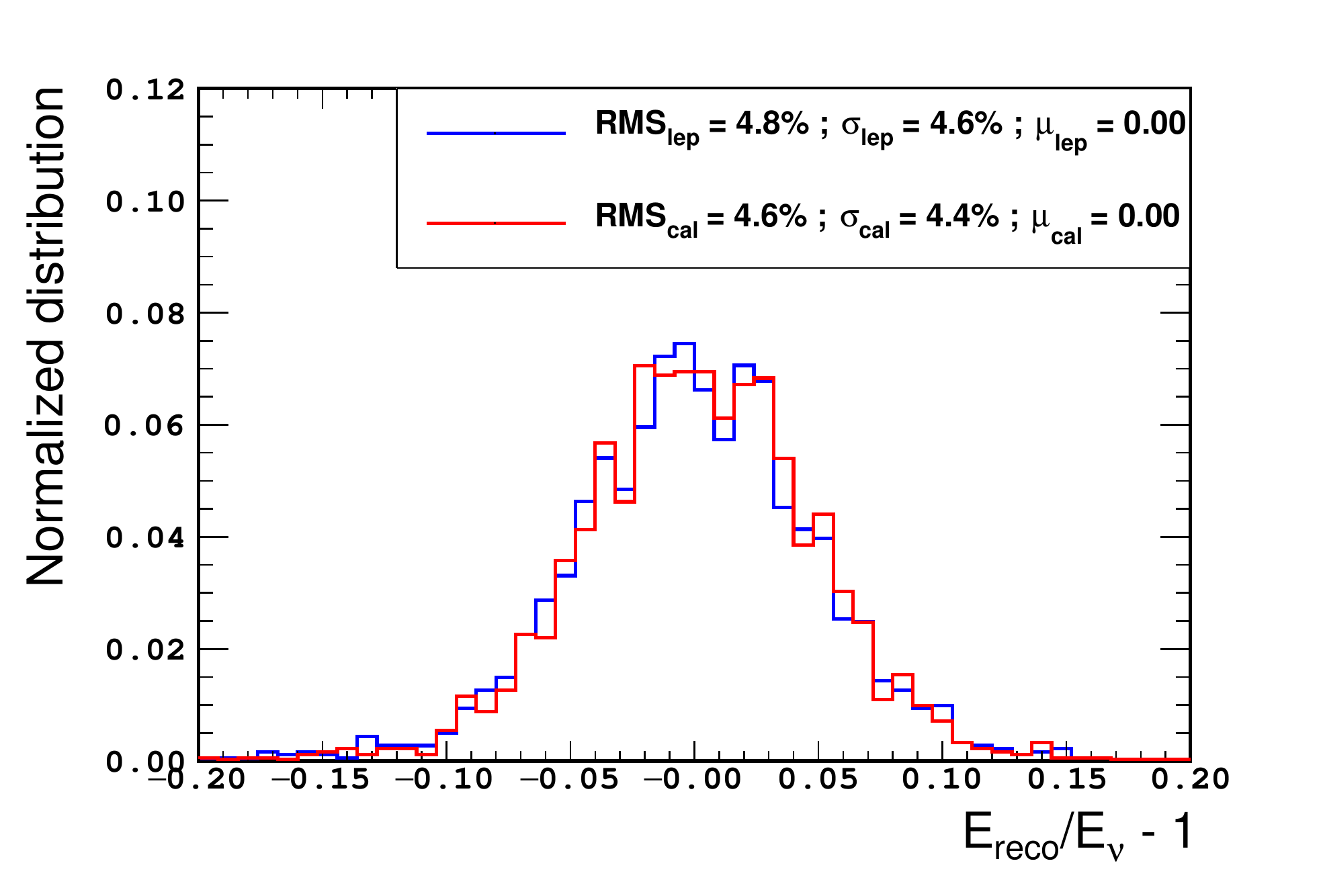}
    \caption{Obtained resolution on the interacting antineutrino with the two different formulas for $\delta p_T < \SI{40}{MeV}$, lever arm $>$ 10cm.}
    \label{fig:finalEnuRes}
\end{figure}

For all the selected events from the reconstructed sample, the antineutrino energy is reconstructed using the two formulas (\ref{eq:Elep}) and (\ref{eq:Ecal}). The result of the antineutrino resolution after applying a cut on $\delta p_T$ is given in Fig.~\ref{fig:finalEnuRes}. It can be seen that both reconstruction methods give a very similar result with an antineutrino energy resolution around 4.5\%.

\begin{figure}[ht]
    \includegraphics[width=\linewidth]{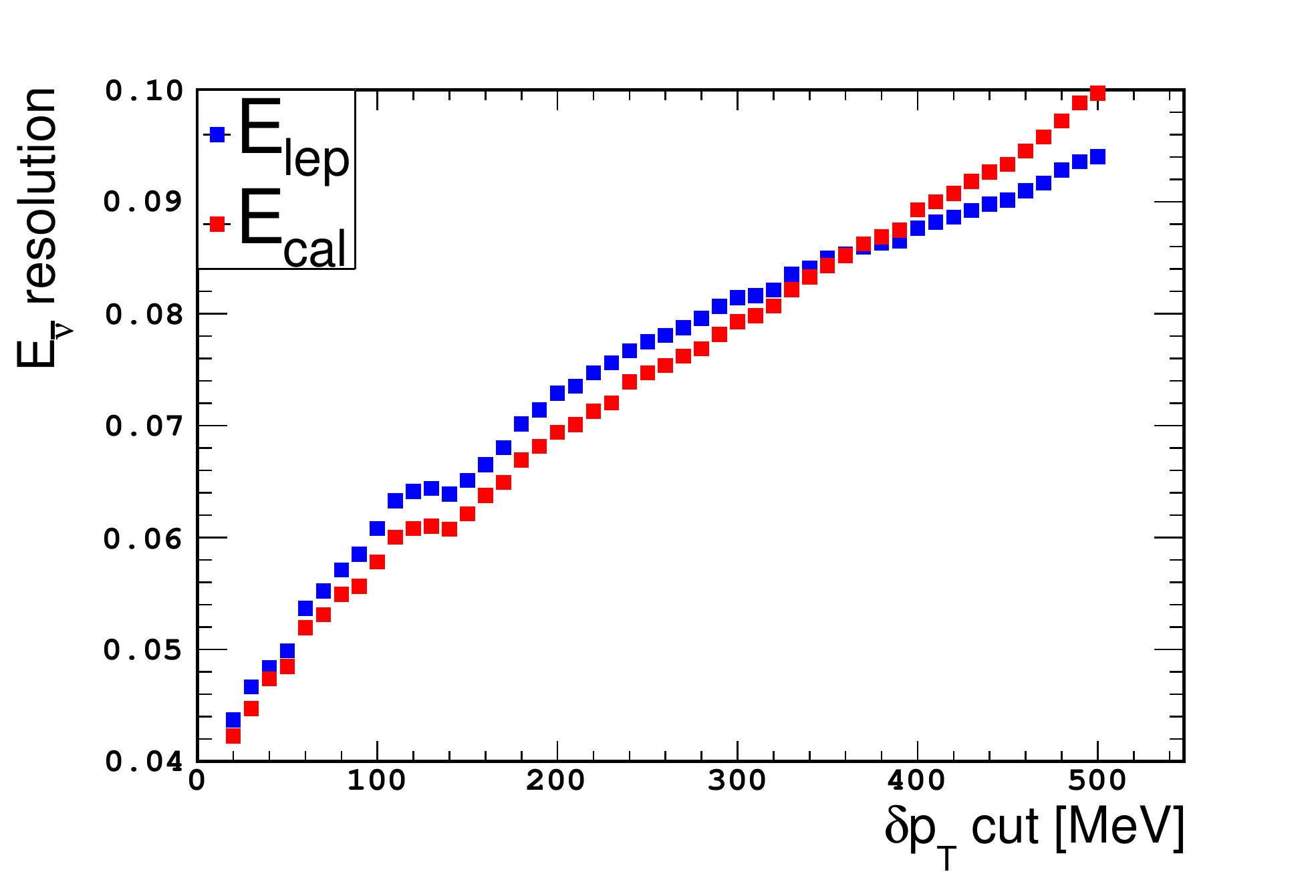}
    \caption{Evolution of the resolution on the reconstructed anti neutrino energy as a function of the $\delta p_T$ cut for \SI{0.5}{\nano\second} time resolution, lever arm $>$ 10 cm.}
    \label{fig:Enu_res_dpt-50}
\end{figure}

\begin{figure}[ht]
    \includegraphics[width=\linewidth]{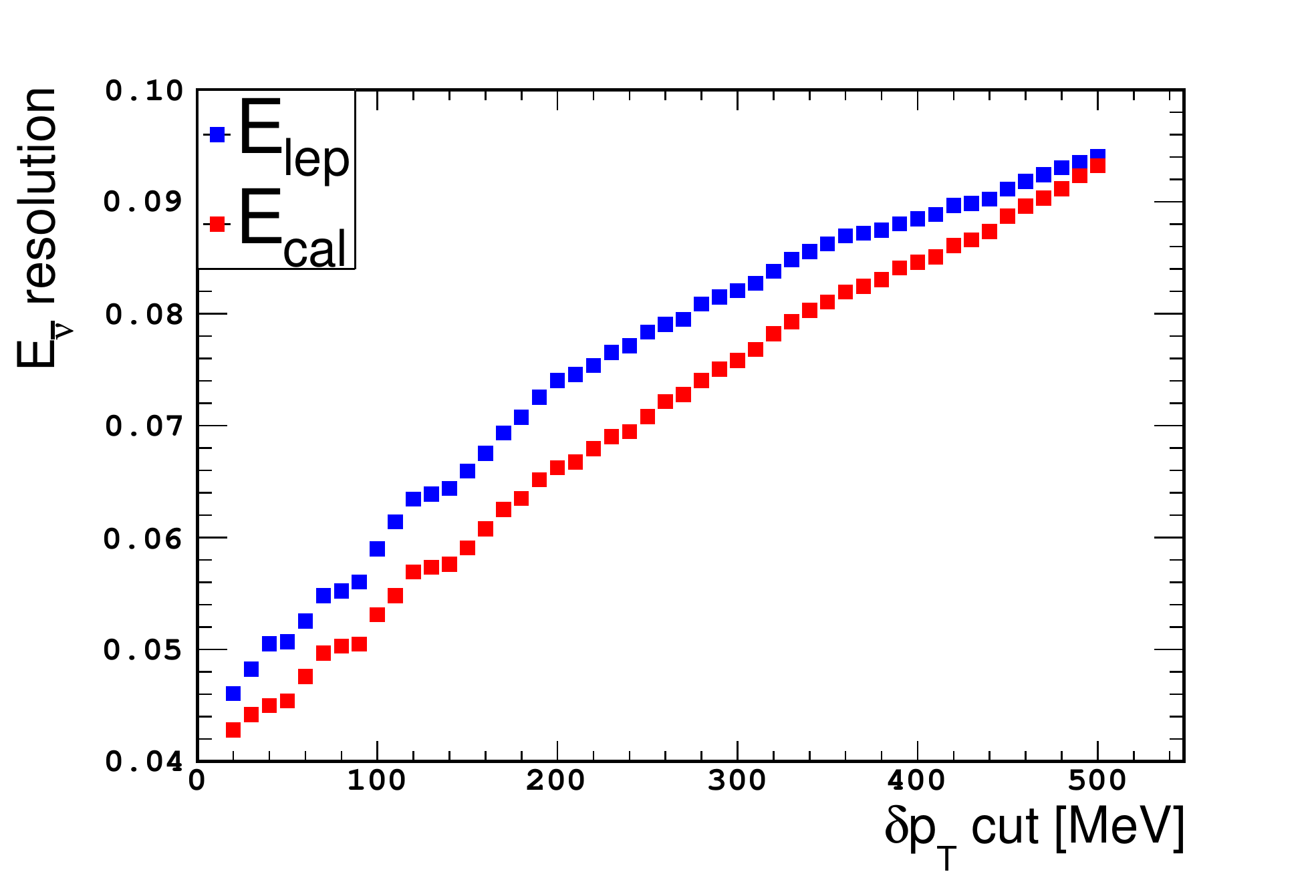}
    \caption{Evolution of the resolution on the reconstructed anti neutrino energy as a function of the $\delta p_T$ cut for \SI{0.25}{\nano\second} time resolution, lever arm $>$ 10 cm.}
    \label{fig:Enu_res_dpt-25}
\end{figure}

\begin{figure}[ht]
    \includegraphics[width=\linewidth]{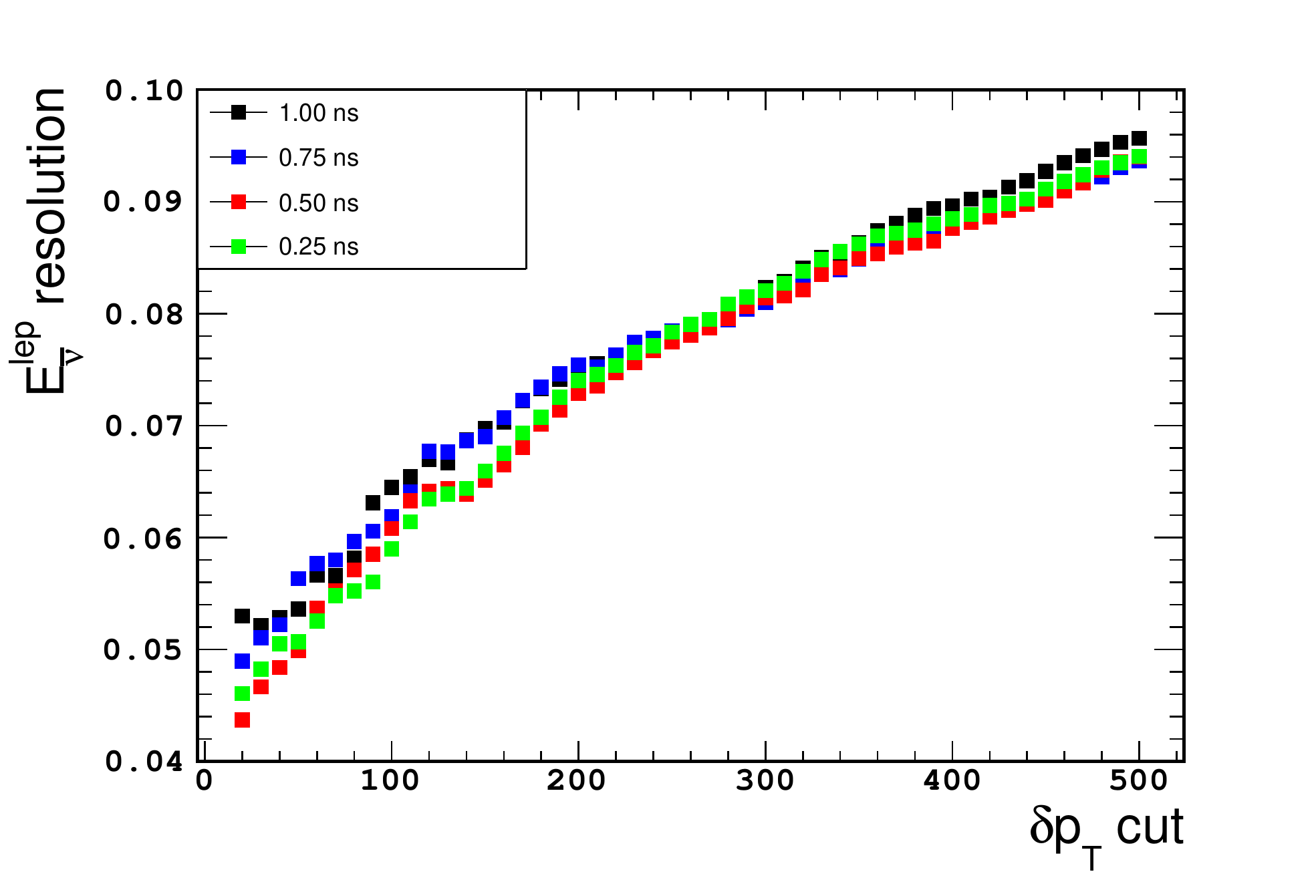}
    \caption{Evolution of the resolution on the reconstructed anti neutrino energy as a function of the $\delta p_T$ cut for different time resolutions for $E_{\bar{\nu}}^{\text{lep}}$.}
    \label{fig:Enu_res-lep}
\end{figure}

\begin{figure}[ht]
    \includegraphics[width=\linewidth]{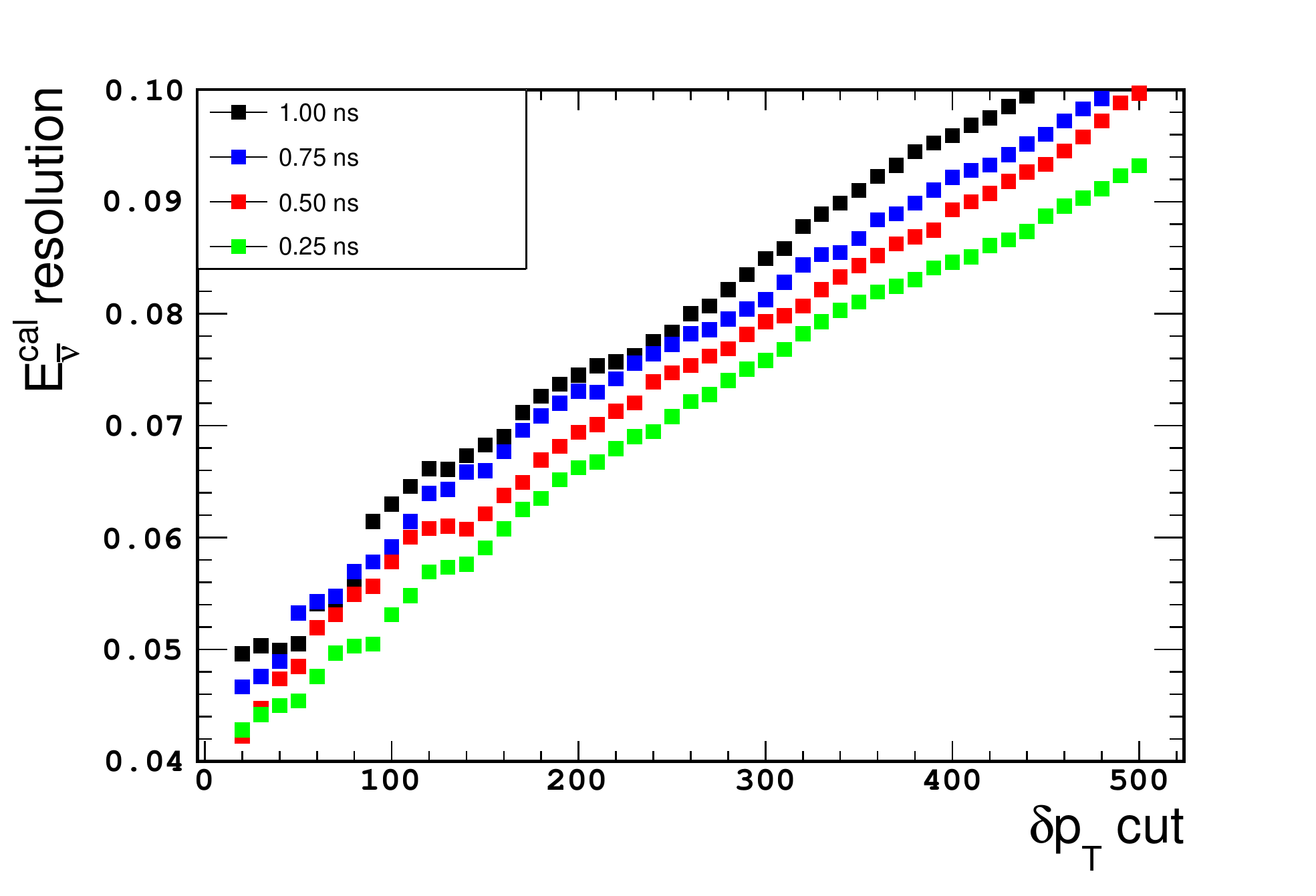}
    \caption{Evolution of the resolution on the reconstructed anti neutrino energy as a function of the $\delta p_T$ cut for different time resolutions for $E_{\bar{\nu}}^{\text{cal}}$.}
    \label{fig:Enu_res-cal}
\end{figure}

Furthermore, the impact on the neutrino energy resolution of the $\delta p_T$ cut and of the detector time resolution is presented in Fig\textcolor{black}{s}.~\ref{fig:Enu_res_dpt-50}-\ref{fig:Enu_res-cal}. Fig.~\ref{fig:Enu_res_dpt-50} shows that for a time resolution of \SI{0.5}{\nano\second}, the two estimations of the neutrino energy have similar performance. Moreover, it can be seen that imposing stricter $\delta p_T$ cuts allows for improving the neutrino energy resolution. In addition, improving the time resolution results in a better energy reconstruction using the calorimetric measurement as shown in Fig\textcolor{black}{s}.~\ref{fig:Enu_res_dpt-25} and \ref{fig:Enu_res-cal}. The improvement in energy resolution with time resolution is mainly noticeable for the calorimetric estimation of the energy, while it remains limited with the leptonic-only estimation as shown in Fig.~\ref{fig:Enu_res-lep}. The time resolution directly impacts the uncertainty on the neutron time-of-flight that is used to estimate its kinetic energy.


\begin{figure}
    \centering
    \includegraphics[width=\linewidth]{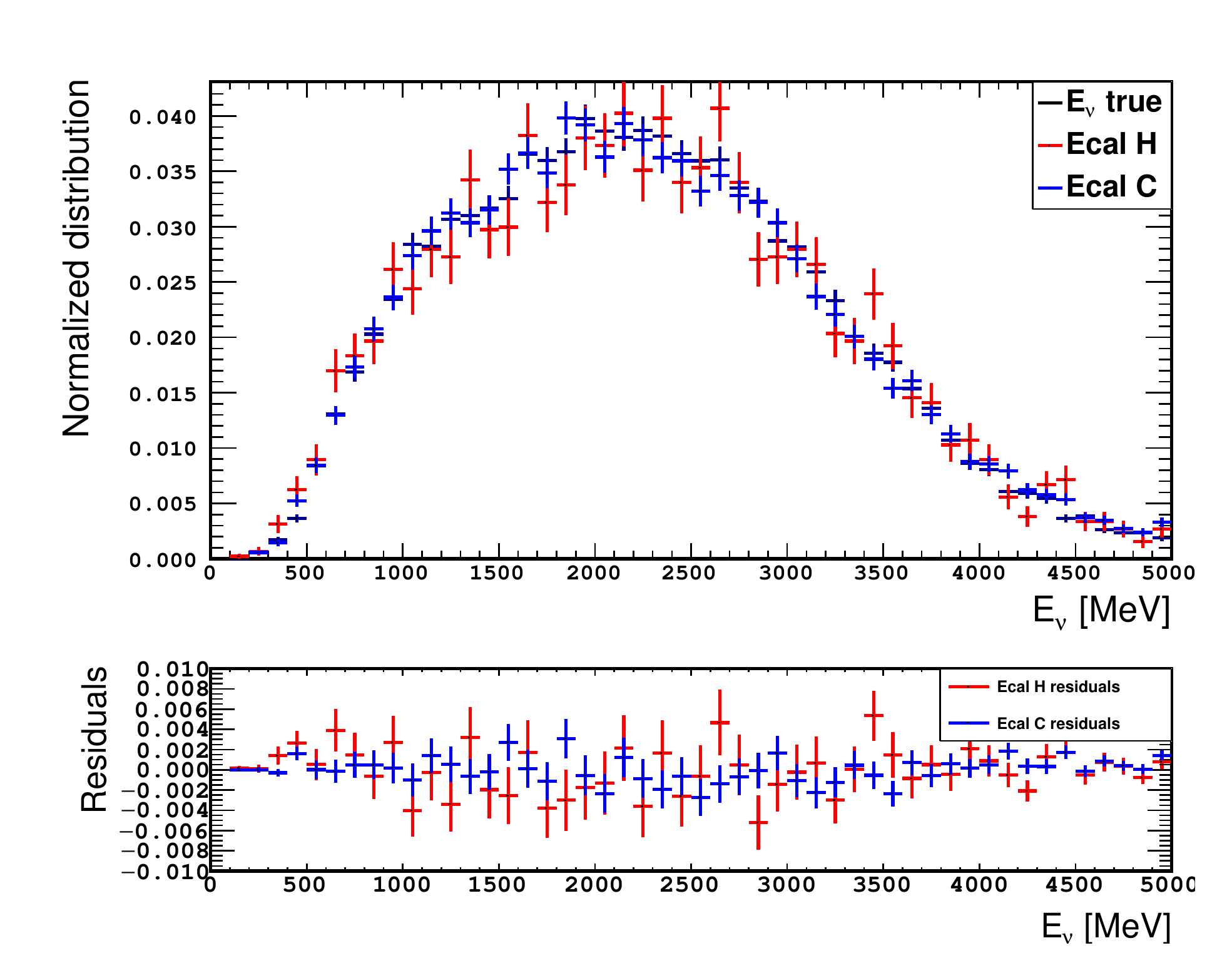}
    \caption{Reconstructed neutrino spectra and difference with the true one, lever arm $>$ 10 cm.}
    \label{fig:recoDistortions}
\end{figure}

The energy resolution is not the only metric to assess the 3DST performance. It is also necessary to check that the reconstructed neutrino energy spectrum is not distorted, given such a detector to be installed as a near detector for a long-baseline neutrino oscillation experiment.
Fig.~\ref{fig:recoDistortions} shows that the energy reconstruction presented here does not distort the reconstructed neutrino energy spectrum, and either a selection of H or C interactions has no impact on the shape of the reconstructed spectrum with $\chi^2$ test p-values above 0.2 for both cases ($\chi^2/d.o.f.$ = 58/50).

\begin{figure}
    \centering
    \includegraphics[width=\linewidth]{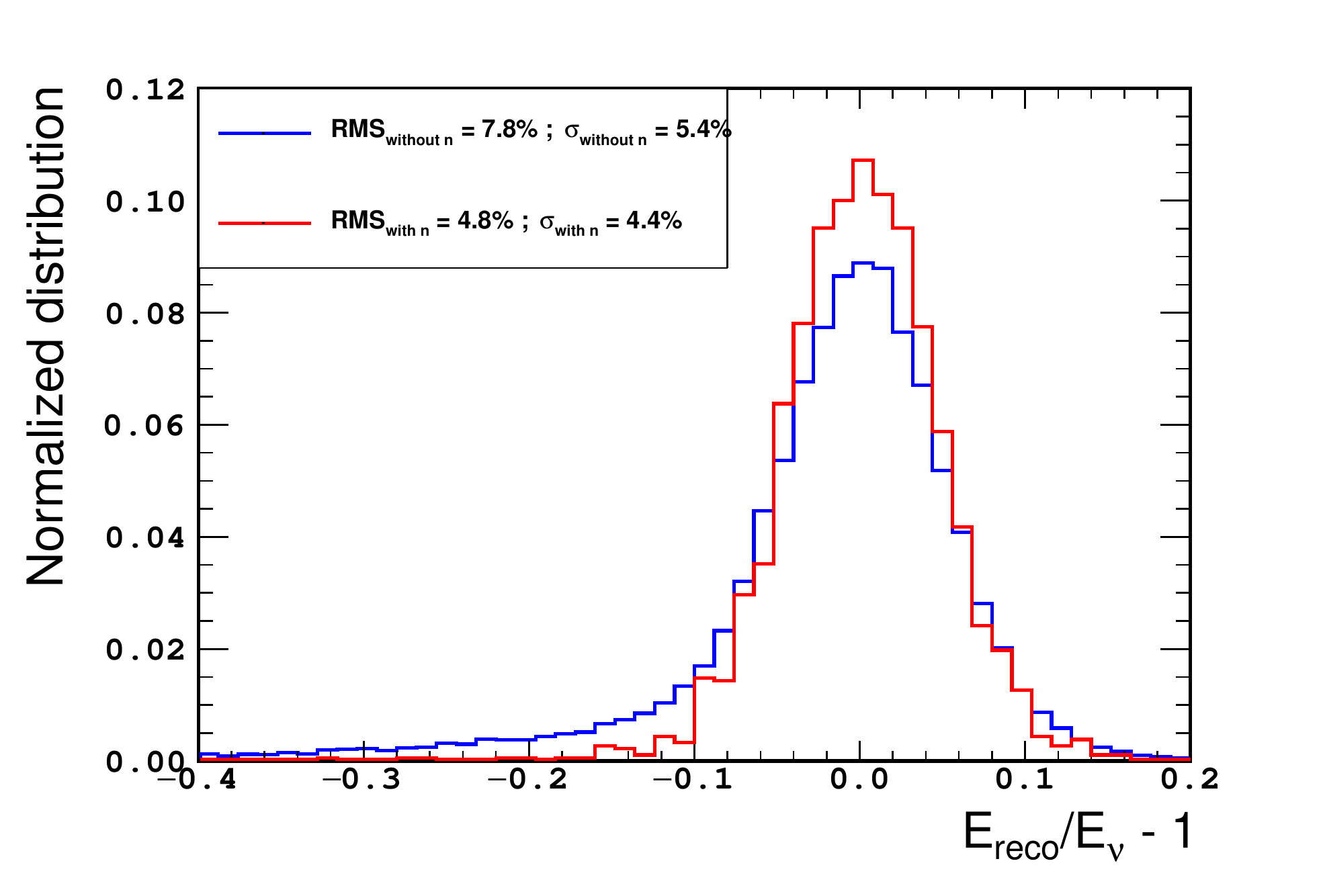}
    \caption{Obtained resolution on the interacting antineutrino with and without the neutron information for a \SI{1}{ns} time resolution. The resolution without the neutron detection is represented by the leptonic energy reconstruction with no $\delta p_T$ cut, while the one with neutron detection uses the calorimetric energy reconstruction and $\delta p_T < \SI{40}{MeV}$.}
    \label{fig:legit}
\end{figure}

Finally, one can fully measure the \textcolor{black}{benefit} of neutron detection on the neutrino energy resolution by comparing the obtained resolutions with and without the detection of the neutrons. Without neutron detection, there is no way to estimate the $\delta p_T$. This is reflected in Figure~\ref{fig:legit} where the neutrino energy resolution obtained using $E_\text{lep}$ without any $\delta p_T$ selection is compared to the one obtained using the neutron information fully. It can be seen that the neutron detection capabilities of such a detector allow for a substantial improvement of the neutrino energy measurement even for the \SI{1}{ns} conservative time resolution. A significant part of this improvement can be found in the suppression of the lower-end tail, corresponding to an underestimation of the neutrino energy, mostly due to the availability of the $\delta p_T$ measurement.

\section{CC0\texorpdfstring{$\pi$}{Lg}0p1n channel analysis with neutron}
\label{section6}
On the top of the CC0$\pi$ selection in the previous section, the $\bar{\nu}_\mu$ CC0$\pi$0p1n channel is further studied to constrain the flux given its relatively small uncertainty on the cross section and detection.
Due to the neutron detection capability, the 3DST can constrain $\bar{\nu}_\mu$ flux with the neutron, analogous to the current detectors constraining ${\nu}_\mu$ flux with a proton in the final state.
This section describes the selection of the CC0$\pi$0p1n channel and further investigates the $\bar{\nu}_\mu$ flux constraint.

\subsection{Neutron selection}
\label{subsec:neutron_selection}
As mentioned in Chapter~\ref{sec:neutron detection performance}, a CC0$\pi$0p sample can be selected and denoted as the single-track sample.
We can select a neutron sample among the single-track sample.
The neutron selection will be described later.
The neutron sample may contain three kinds of backgrounds:
\begin{itemize}
    \item External background: the first object comes from an external source, such as the neutrons from a neutrino interaction outside of the detector fiducial volume.
    \item Internal non-neutron background: the first object comes from the targeted neutrino interaction, but it is not neutron-induced such as \textcolor{black}{a} $\pi^0$ from the neutrino interaction.
    \item Internal neutron background: 
    by design, we are capable of detecting only one neutron kinematics. \textcolor{black}{A} ``multi-neutron event'' is defined as \textcolor{black}{an} event with more than one neutron in the final state. In multi-neutron events, other neutrons are missed, causing the misreconstruction of the neutrino energy.
\end{itemize}
The external background can be reduced to 1\% with cuts on the time difference and distance between the neutrino interaction vertex and the neutron-induced object. The purity as a function of the time difference and lever arm for excluding the external background can be found in Fig.~ 137 in \cite{ND_CDR}.

One of the main sources of the internal non-neutron background is delta rays induced by the primary muon track.
In order to reject them, objects inside a cylindrical region with a radius of 4.25 cm surrounding the muon track are removed, as shown in Fig.~\ref{fig:isolated_object}.

\begin{figure}[h!]
\includegraphics[scale=0.45]{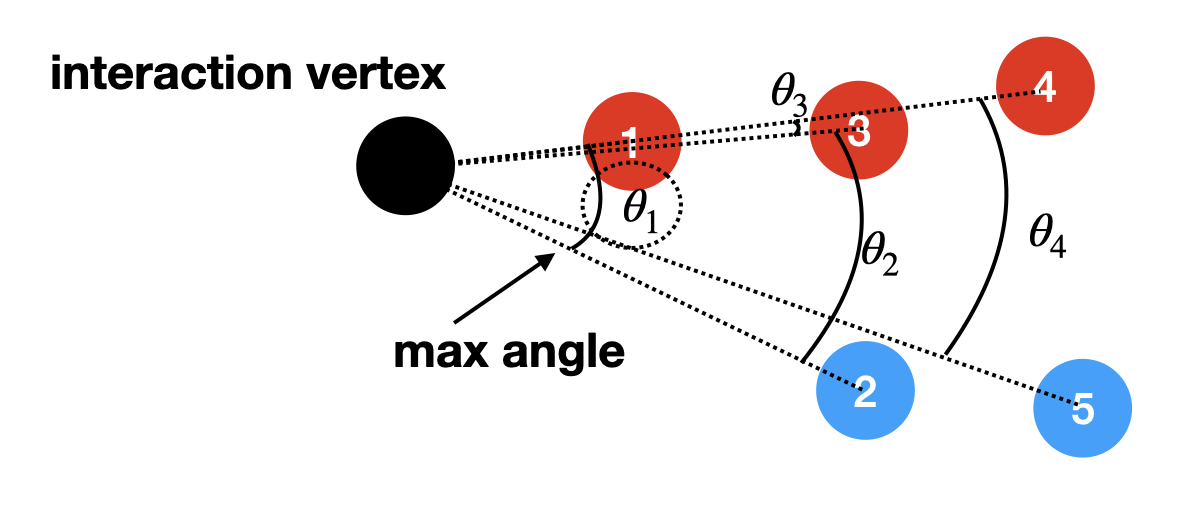}
\caption{The maximum angle in the two-neutron event. Each color represents a list of objects induced by one neutron, and the numbers follow the time order. The angles can be obtained between two adjacent objects, and the biggest one is defined as a ``maximum angle''. The ``maximum distance'' can be defined in a similar sense.}
\label{fig:max_angle_definition}
\end{figure}

The other source of the internal non-neutron background is \textcolor{black}{a} $\pi^0$ from the neutrino interaction vertex.
There is also a small amount of deexcitation photons in the neutrino interaction.
To reduce these, the following cuts are applied to the first object in time.
\begin{itemize}
\item ToF: negative ToF events are rejected to reduce misreconstructed events due to the timing resolution of the detector.
\item Energy deposit: the total energy deposit of the neutron-induced object tends to be higher than others.
\item Branch number: an object can induce small tracks looking like branches attaching to the object. For neutron-induced object, the branch number, defined as the number of small tracks, tends to be lower since neutron mainly produces visible single-track protons. 
\end{itemize}
The distributions for those variables above and the value of the cuts can be found in Appendix~\ref{Appen:A}.
At this stage, the selected sample has a 90\% purity of neutron candidates with 49\% efficiency. Additional selections are needed to reduce the internal neutron background.
Multi-neutron events can have a large spread of isolated neutron-induced objects in the plane transverse to the incident neutrino compared to single-neutron events, as shown in Fig.~\ref{fig:max_angle_definition}.
In order to reduce multi-neutron events, the angles and the distances between the adjacent objects in time are measured.
Then we pick the biggest angle and distance among the isolated objects as the  ``maximum angle'' and ``maximum distance'' respectively.
Fig.~\ref{fig:max_angle_distance} and~\ref{fig:max_distance_dist} shows the distributions of the maximum angle and distance. The events with values smaller than the cuts are selected.

Lastly, if the primary neutron interacts in the detector without leaving enough energy and interacts again with a high enough energy deposit, the first scattering is not visible. This invisible scattering is not a major background, given that the invisible scattering is mostly elastic, and it does not change the neutron angle significantly.

\begin{figure}
\centering
\includegraphics[scale=0.45]{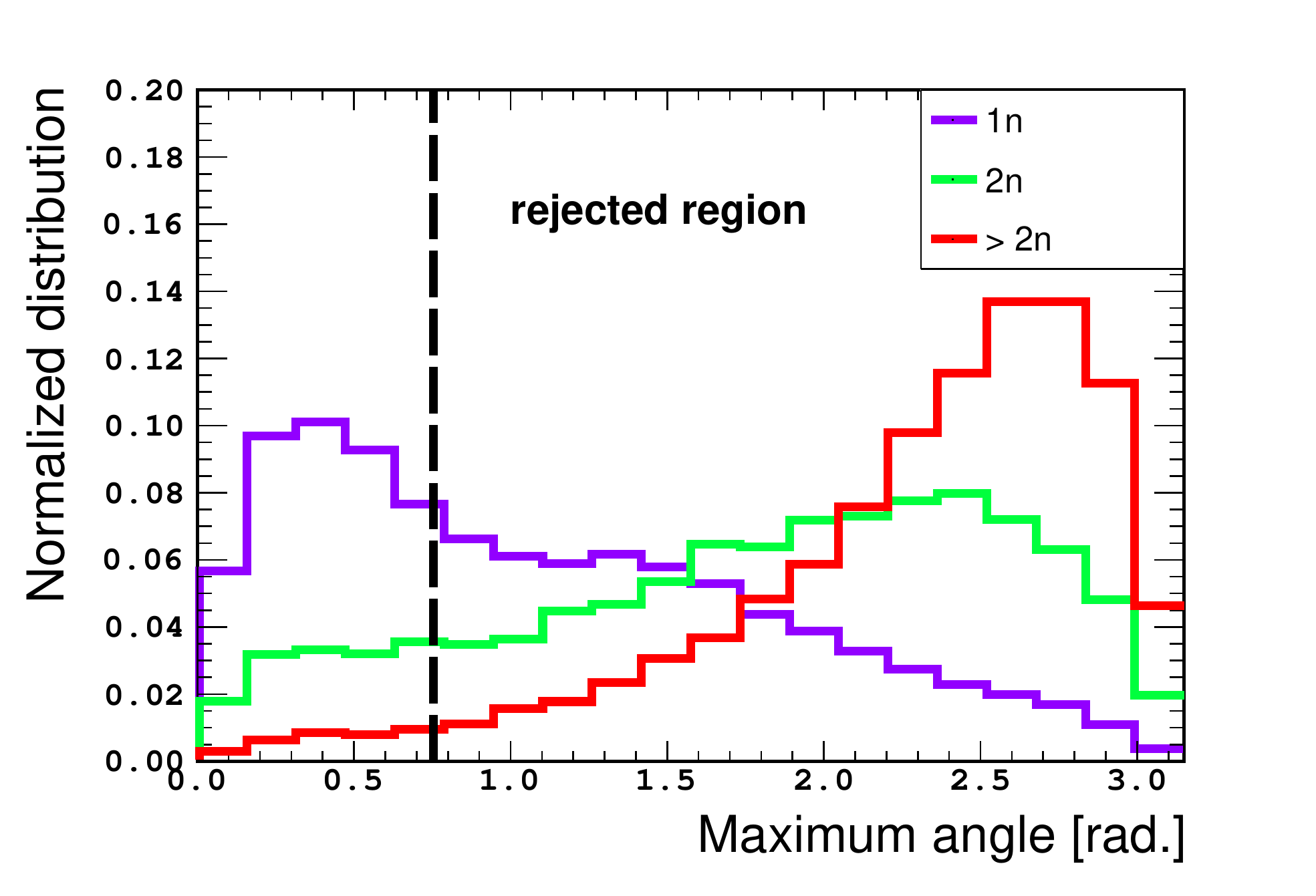}
\caption{Maximum angle for various cases of neutron multiplicity. It shows the separation of single-neutron and multiple-neutron cases. The dashed line shows the selection cut.
}
\label{fig:max_angle_distance}
\end{figure}
\begin{figure}
\centering
\includegraphics[scale=0.45]{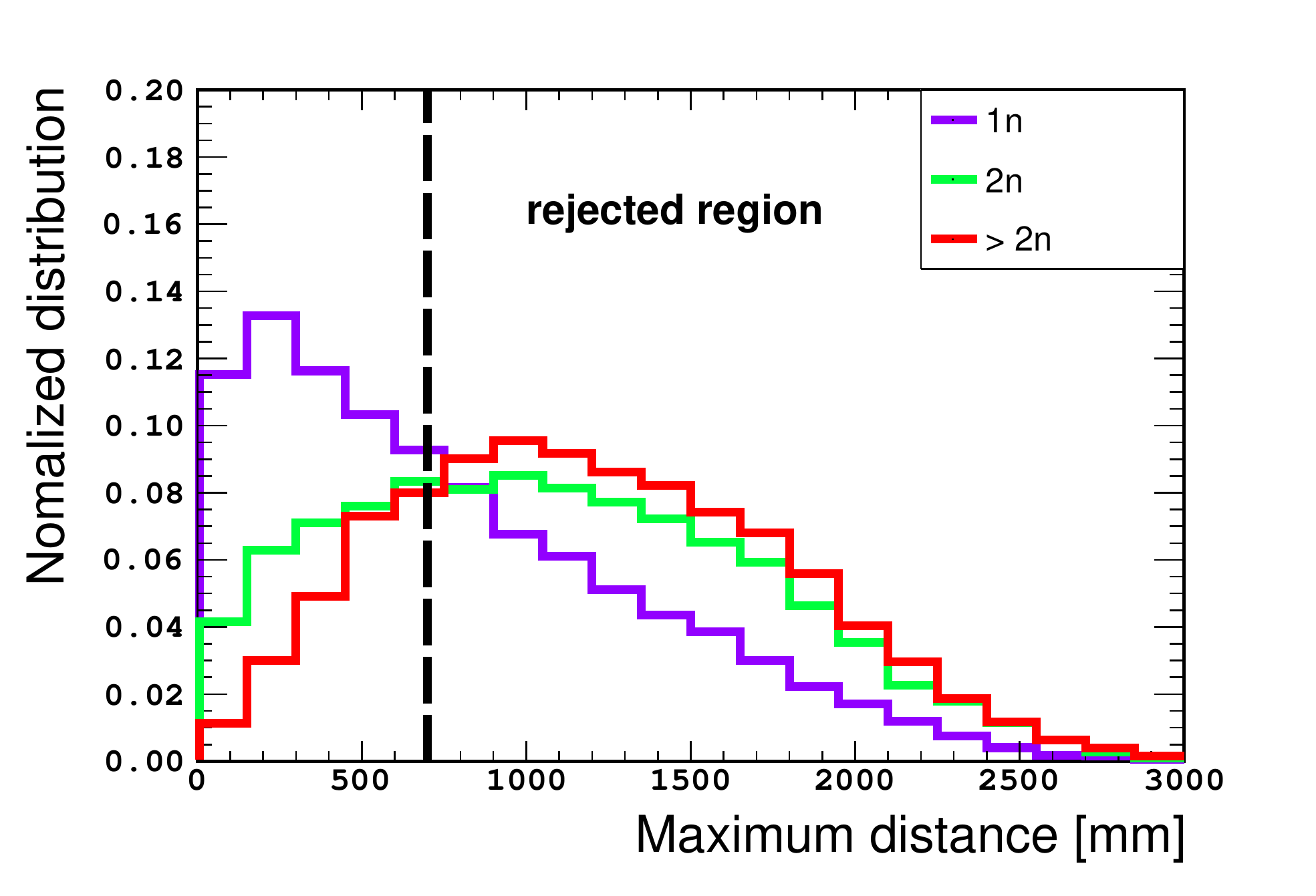}
\caption{Maximum distance for various cases of neutron multiplicity. It shows the separation of single-neutron and multiple-neutron cases. The dashed line shows the selection cut.
}
\label{fig:max_angle_distance}
\end{figure}

\subsection{Efficiency and purity}
With all the selection cuts presented in previous section, a significant background reduction is achieved.
\begin{table}[h!]
\centering
\begin{tabular}{ |p{4cm}||p{1.5cm}|p{1.5cm}|  }
 \hline
 \multicolumn{3}{|c|}{Purity and efficiency} \\
 \hline
 Cut& purity & efficiency \\
 \hline
  ToF~(including threshold)  &   0.48  & 0.70\\
 energy deposit &   0.50  & 0.58\\
 branch number & 0.53 & 0.54\\
 max angle    & 0.80 & 0.27\\
 max distance & 0.81 & 0.23\\
 \hline
\end{tabular}
\caption{Purity and efficiency for each step of selection. The selections are applied step by step to the sample. The purity is the number of signal samples divided by the number of samples after the cuts, and the efficiency is the number of samples after the cuts divided by the number of samples before the cuts. }
\label{table:1}
\end{table}
Table.~\ref{table:1} shows the purity and efficiency of the sample at each selection step.
There is a significant reduction of efficiency by the energy deposit cut. In the neutron beam test, we realized \textcolor{black}{that there was} a non-negligible amount of electronic noise and cross-talk light through the cube holes. The energy deposit cut can efficiently remove almost all the noise and cross-talk light.
In the end, the signal samples, CC0$\pi$0p1n, have a purity and efficiency of 81\% and 23\%.
\begin{figure}[h!]
\includegraphics[scale=0.45]{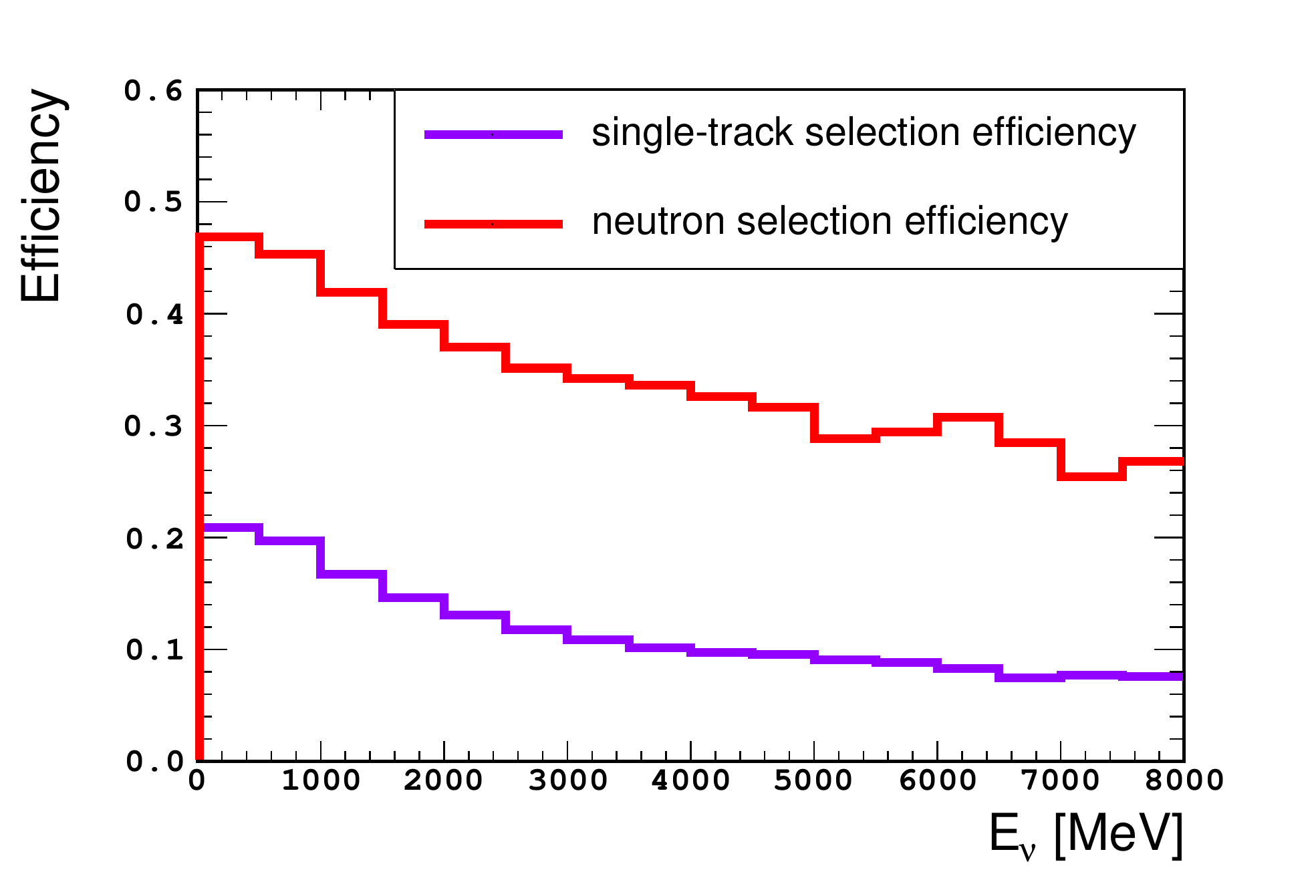}
\caption{The efficiency curves of each selection group which are single-track and neutron selection.}
\label{fig:effCurves}
\end{figure}
Fig.~\ref{fig:effCurves} shows the efficiency as a function of E$_\nu$ for the CC0$\pi$0p selection and the neutron selection.
Note that each efficiency has a different denominator. 
In this analysis, we assume the efficiency uncertainty can be well measured across the energy distribution and is ignored.

\subsection{CC0\texorpdfstring{$\pi$}{Lg}0p1n fitting}
\label{subsec:fitting}
With the selected CC0$\pi$0p1n sample, a sensitivity study is performed to investigate the capability of constraining DUNE flux uncertainties.
There are 256 parameters used to account for the various systematic uncertainties of the flux, such as hadron production, beam focusing mode, horn alignment, etc~\cite{DUNEflux}.
A principal component analysis (PCA) is used to obtain the 1$\sigma$ uncertainty as a function of true E$_\nu$~\cite{DUNE_TDR}.
\begin{figure}
\includegraphics[scale=0.45]{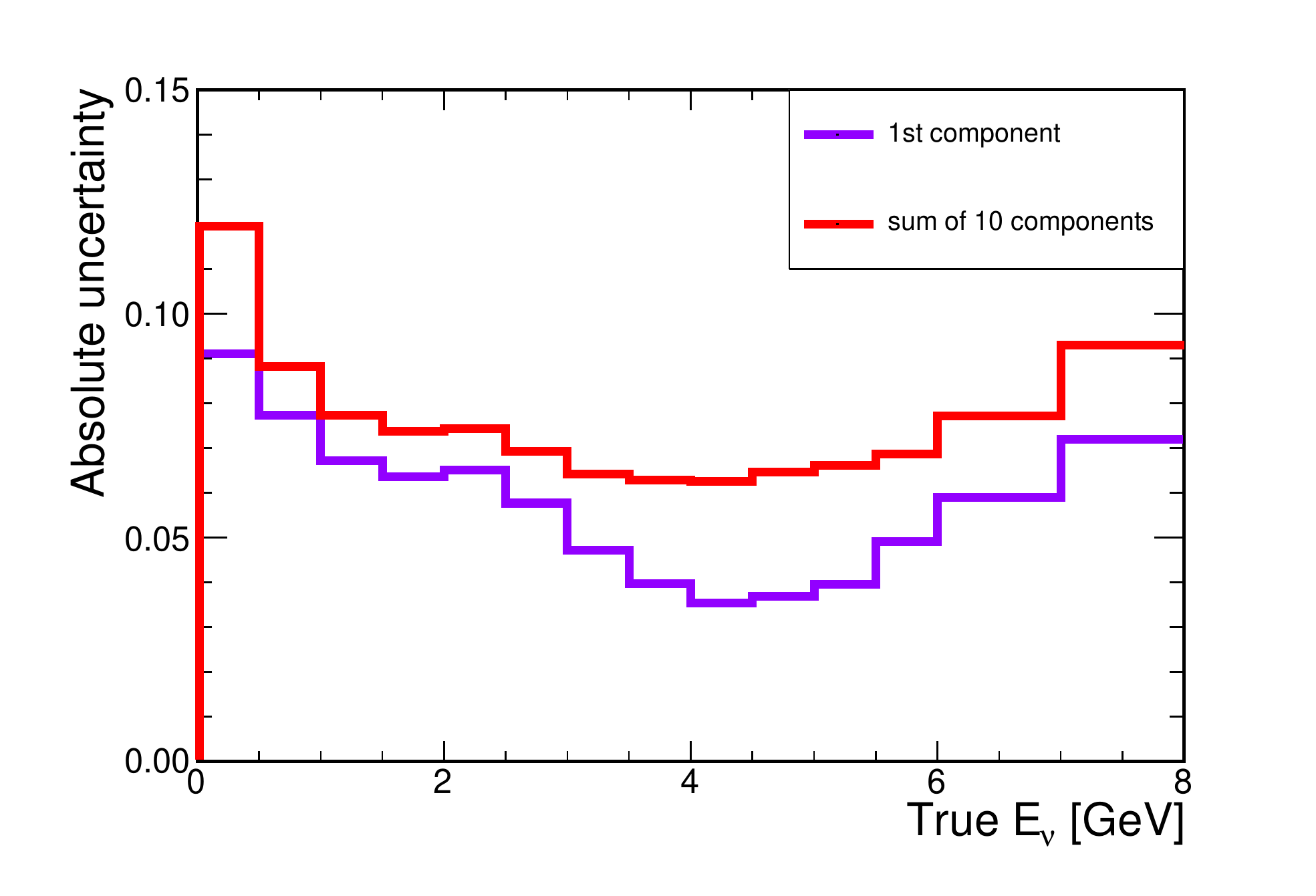}
\caption{The absolute value of the flux systematic uncertainties. 
The biggest component (hadron production) and the sum of the biggest 10 of them are shown in this figure. 
The sum will be used as a pre-fit uncertainty.}
\label{fig:flux_systematics}
\end{figure}
The biggest 10 of them, covering 95\% of the variance, are used in this analysis, and the largest one and the sum of the ten are shown in Fig.~\ref{fig:flux_systematics}.

A $\chi^2$ fitting framework is developed for this study with $\chi^2$ defined as
\begin{equation}
\begin{split}
\chi^2 = \sum(P-D)^T*M_{cov}^{-1}*(P-D)  \\
+ \sum^{10}_{i = 1} \frac{\left( f_{i,CV} - f_i \right)^2}{\sigma_{f_i}} 
+ \frac{(f_{B,CV} - f_B)^2}{\sigma_{f_B}}  \\
+ \frac{(f_{e,CV} - f_e)^2}{\sigma_{f_e}},
\label{eq:chi2}
\end{split}
\end{equation}
where $f$ is the pull term with \textcolor{black}{the} subscripts $i$, $B$, and $e$ indicating flux, background, and energy scale, respectively.
The $D$ is the CC0$\pi$0p1n fake data sample. 
The $M_{cov}$ is a covariance matrix that includes the statistical uncertainty and the cross-section uncertainty. 
The $CV$ is the central value, which is set to 0, and the $\sigma_{f}$ is set to 1.
The $P$ is the predicted energy spectrum reweightable by $w$ and the energy scale, defined as
\begin{equation}
\begin{split}
P = (P_0 \times w) \times (1 + E_{scale} \times f_e),~\text{and}\\
w = \prod^{10}_{i=0} \bigg( 1 + f_i \times syst_i \bigg)\text{.}
\end{split}
\end{equation}
The $E_{scale}$ is the 1 $\sigma$ shift on the energy spectrum due to the energy scale uncertainty.
The $syst_i$ is the  1 $\sigma$ shift on the energy spectrum for the $i^{\text{th}}$ flux PCA component.
The $P_0$ is the nominal predicted energy spectrum.

In summary, the following systematic uncertainties are considered in the fitting framework: 
\begin{itemize}
\item DUNE flux systematic uncertainty;
\item Cross section uncertainty: The GENIE Reweight package is used~\cite{genie_reweight}. GENIE has a list of cross-section parameters. All those parameters are varied simultaneously 1,000 times to extract the integrated cross-section uncertainty as a function of the neutrino energy. There is no correlation assumed among the cross-section parameters. The correlations among neutrino energy bins are embedded in the parameter variations according to GENIE.
In addition, the bias caused by the generators is considered.
The default neutrino interaction generator is GENIEv3. The GENIEv2 has a largest discrepancy from the GENIEv3~\cite{https://doi.org/10.48550/arxiv.2203.11821}.
\textcolor{black}{The cross-section discrepancy can be up to 10\%, depending on the neutrino energy. The largest discrepancy appears at the neutrino energy below 1 GeV, and the discrepancy decreases as the neutrino energy increases.}
The difference between them is taken as an additional cross-section modeling uncertainty;
\item Background uncertainty: the background uncertainty is assumed to be 100\%, and it acts as an overall normalization shift.
\item Energy scale: the neutrino energy is varied with smearings of neutron energy and $\mu^+$ energy by 20\% and 2\%, respectively. 
The 1$\sigma$ from the resulting Gaussian distribution of the neutrino energy is taken as the energy scale uncertainty.
\end{itemize}

\begin{figure}
\centering
\includegraphics[scale=0.45]{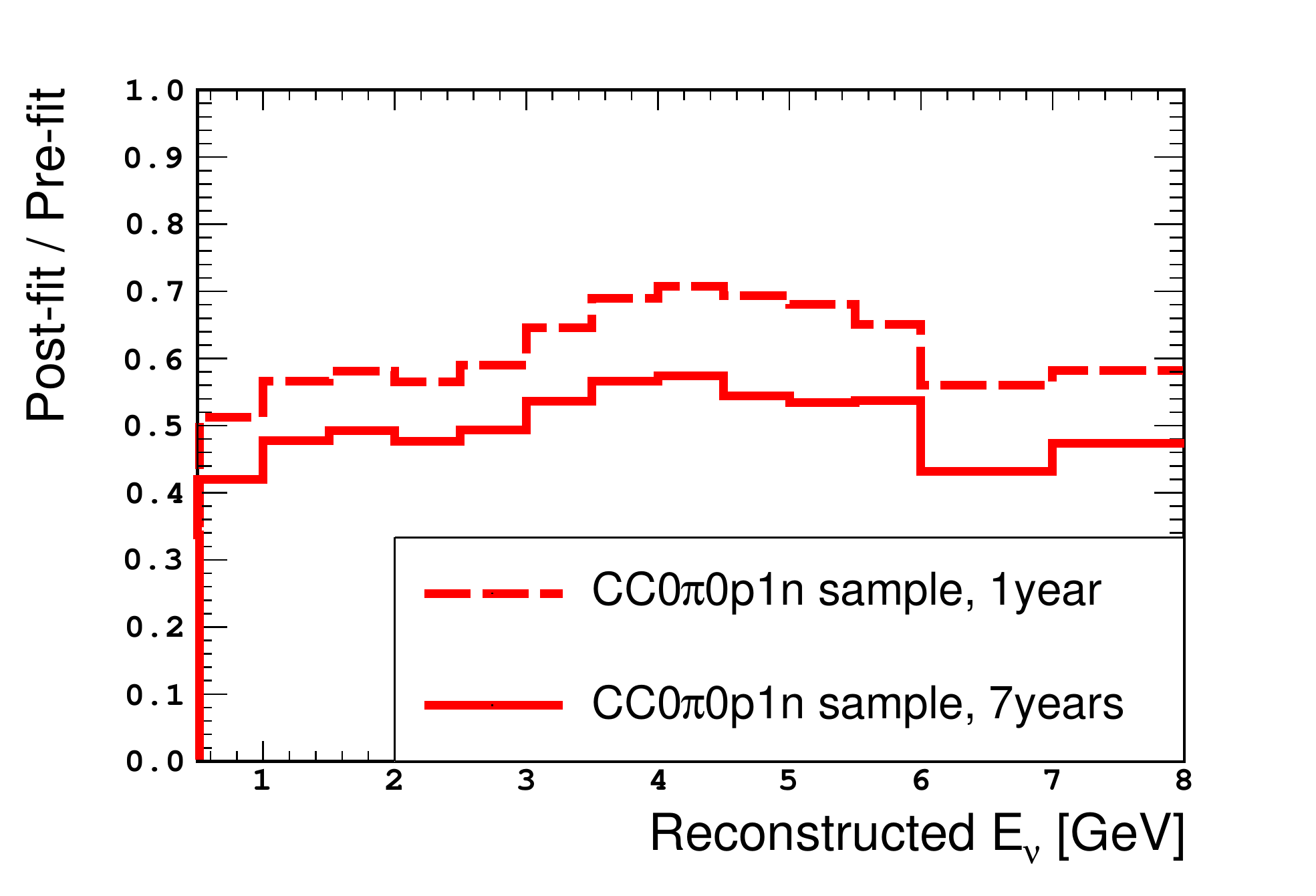}
\caption{Fitting result for flux uncertainty \textcolor{black}{of $\bar{\nu}_{\mu}$ flux in RHC mode} with CC0$\pi$0p1n sample. The ratio of the quadrature sum of the flux uncertainty after and before the fit is presented.}
\label{fig:lownu_result_only_neutron}
\end{figure}

The flux uncertainty constraint with the CC0$\pi$0p1n sample is presented by comparing the post-fit and pre-fit flux uncertainties. The ratio of the post-fit and pre-fit uncertainties is as shown in Fig.~\ref{fig:lownu_result_only_neutron}.  
The statistics are assumed to be with one year and seven years of run time.

\textcolor{black}{
In order to understand the impact of neutron ToF detection, we compare our nominal result with a mock data set with 100\% neutron energy uncertainty. The overall post-fit to pre-fit ratio is around 0.85 with one year run time. The precise neutron energy measurement improves the flux constraint significantly.
An additional test was done to understand the ability of the fitter to recover biased flux prediction. The mock data was tweaked by 1 $\sigma$ bias for the largest flux component. Given the prior pull on the tweaked dial, the fitter can recover the bias by 0.6 $\sigma$, and the remaining discrepancy was recovered by moving other systematic pulls. If there is no prior pull on the tweaked dial, the fitter can fully recover the bias.
}

Furthermore, the MINER$\nu$A experiment demonstrated an in-situ measurement of the CC cross section in the NuMI beamline with the flux prediction obtained by the low-$\nu$ method~\cite{minerva_lownu}.
Following the same idea, the low-$\nu$ method can be used with 3DST as well. More detail is discussed in Appendix~\ref{sec:low_nu}.

\section{conclusion}
We studied the 10-ton-scale 3D\textcolor{black}{-}projection scintillator tracker's capability of detecting neutron kinematics on an event-by-event basis with a full reconstruction and a GeV-scale neutrino beam. The neutron detection precision was presented in detail. Overall neutron energy resolution below 20\% can be achieved with sub-ns timing resolution in the detector. Furthermore, we studied how neutron kinematic detection improves the neutrino energy reconstruction. In particular, the antineutrino energy resolution can go down to a few percent with a transverse momentum cut. In addition, we performed a flux constraint study with the individual neutron selection. The neutron selection purity, efficiency, and potential backgrounds were studied in detail. With a year of exposure, the CC0$\pi$0p1n channel can reduce the flux uncertainty by almost a factor of two.

Event-by-event neutron kinematic detection opens a new era of fully utilizing the final-state particle information in neutrino interactions. The near detector of the next-generation experiments will play a crucial role in understanding the neutrino interaction and neutrino flux at an unprecedented level. Following the neutron detection method in this paper, the next-generation near detectors can use the transverse plane variables to deeply study neutrino-nucleus interactions and constrain the flux. The detector design in this paper is uniquely suited for measuring the transverse momentum variables due to its fast timing, fine granularity, passive material absence, and low threshold. 
The T2K upgrade includes such a detector, and it will lead the exploration of the GeV-scale neutrino interaction. 

On top of the method in this paper, the target-independent neutrino flux measurement can be completed with other complementary methods. 
For example, a $\nu$-$e$ scattering measurement can provide a solid constraint on the neutrino flux with various target materials, including carbon and liquid argon~\cite{nuescatter}. 
Combining neutrino flux constraints in multiple ways can also have a more significant impact.
The constraint with the $\nu$-$e$ scattering method is at a similar level as the CC0$\pi$0p1n method, as indicated in this paper.
The constraint with the CC0$\pi$0p1n sample is from a relatively small uncertainty in the modeling due to simple event topology.
In addition, we are effectively benefitting from the low $\delta p_T$ selection (less nuclear effect) since the neutron is strictly required, thus selecting a relatively low $\delta p_T$ sample. 

It is worth noting that our estimate on the CC0$\pi$0p1n systematic uncertainty predominately relies on the neutrino interaction models, particularly the GENIE and Geant4 models.
Improvement of such models will improve the cross-section uncertainty estimate and make the result more accurate.
Lastly, the current flux uncertainty is limited by knowledge of the hadron production.
In the future, we expect some more precise hadron production measurements from \textcolor{black}{the} NA61/SHINE and EMPHATIC experiments~\cite{NA61, EMPHATIC}.
If we assume a 50\% tighter constraint from the upcoming hadron production experiments, the post-fit to pre-fit ratio will be 0.8 to 0.9 throughout the neutrino energy.

\begin{acknowledgments}
This work was supported by NRF grant funded by MSIT of Korea (NRF-2022R1A2C1009686, NRF-2017R1A2B4004308).
This work was supported in part by the MHES (Russia) grant “Neutrino and astroparticle physics" No. 075-15-2020-778.
We further acknowledge the support of the US Department of Energy, Office of High Energy Physics.
\end{acknowledgments}

\appendix
\section{Variable Distributions}
\label{Appen:A}
The first object in time can be induced by either a neutron or other particles.
Depending on the inducing source, the variable distributions have a distinctive feature. The background can be reduced by a combination of simple 1D cuts \textcolor{black}{on} each variable. \textcolor{black}{There are two types of reconstructed objects: cluster and track.
Depending on the type, there are two distributions of the total energy deposit for the first object.} 
Events with branch number $>$ 0 and energy deposit $<$ \SI{510}{MeV} for cluster case  \textcolor{black}{or} energy deposit $<$ \SI{3600}{MeV} for track case are rejected.

\begin{figure}[h!]
\centering
\includegraphics[scale=0.45]{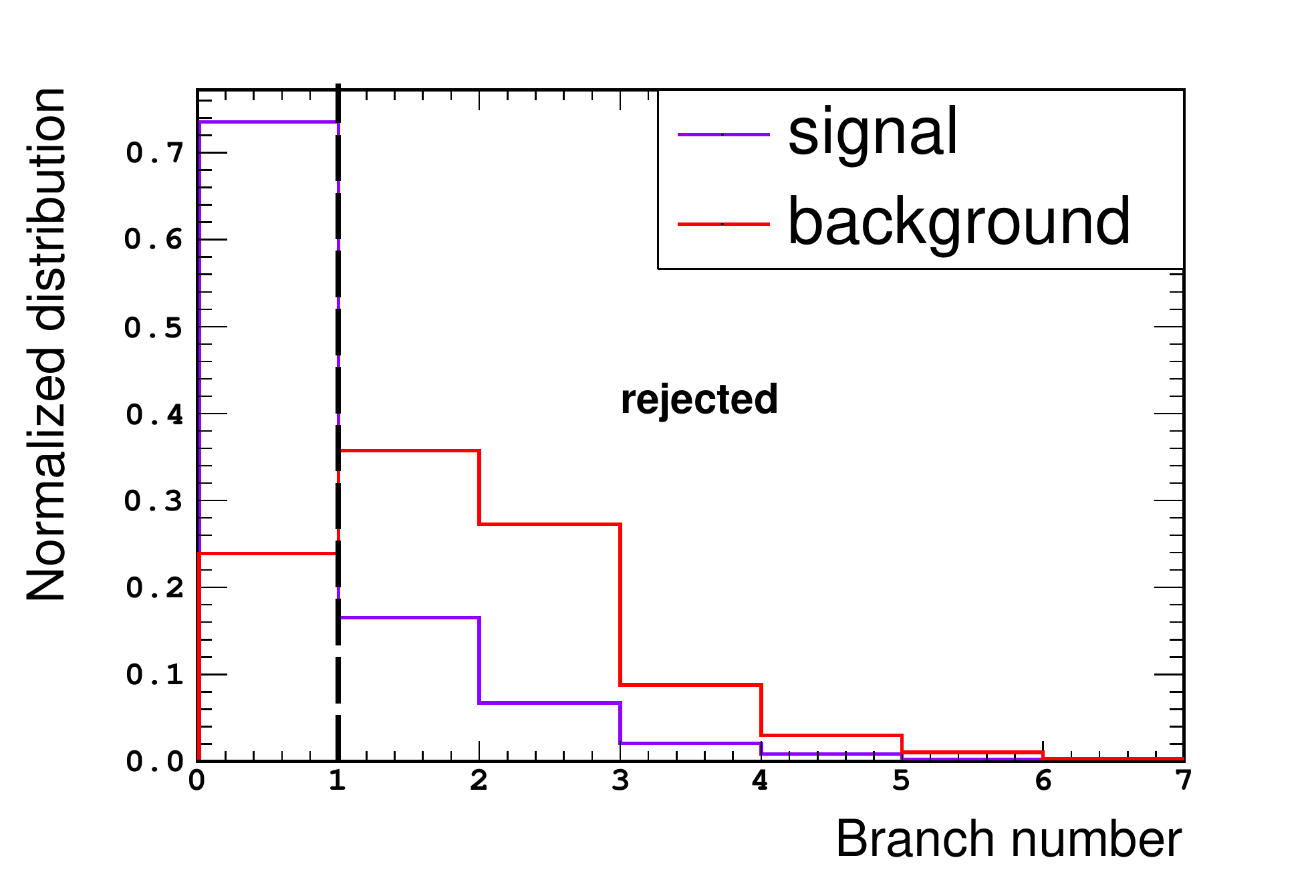}
\caption{Branch number attached to the first object in time. The object can induce particles that look like branches. The signal tends to be lower since neutron interacts less than other particles.}
\label{fig:branch_number}
\end{figure}

\begin{figure}[h!]
\centering
\includegraphics[scale=0.45]{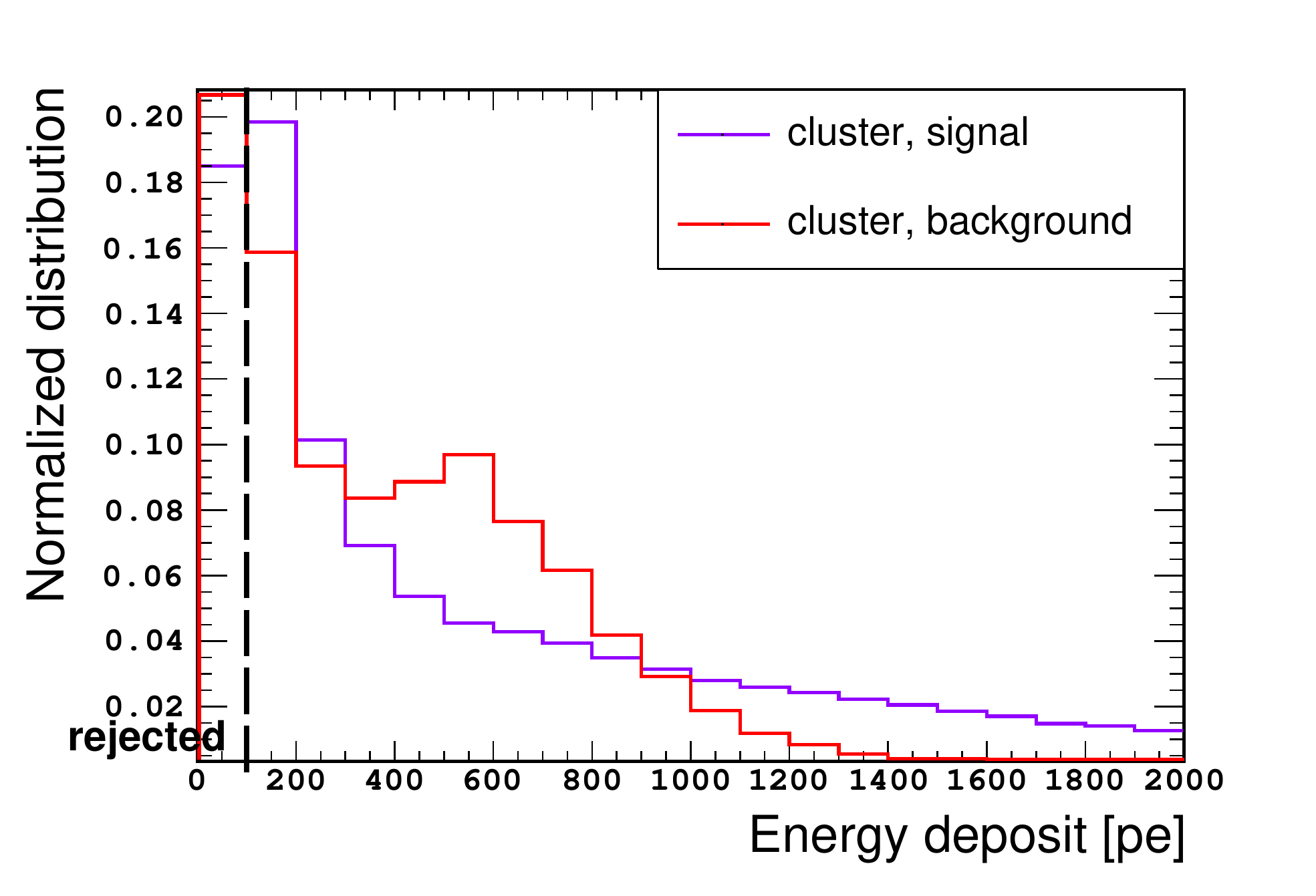}
\caption{The total energy deposit of the first object in time in cluster case. Neutron-induced clusters deposit larger energy.}
\label{fig:edep_cluster}
\end{figure}

\begin{figure}[h!]
\centering
\includegraphics[scale=0.45]{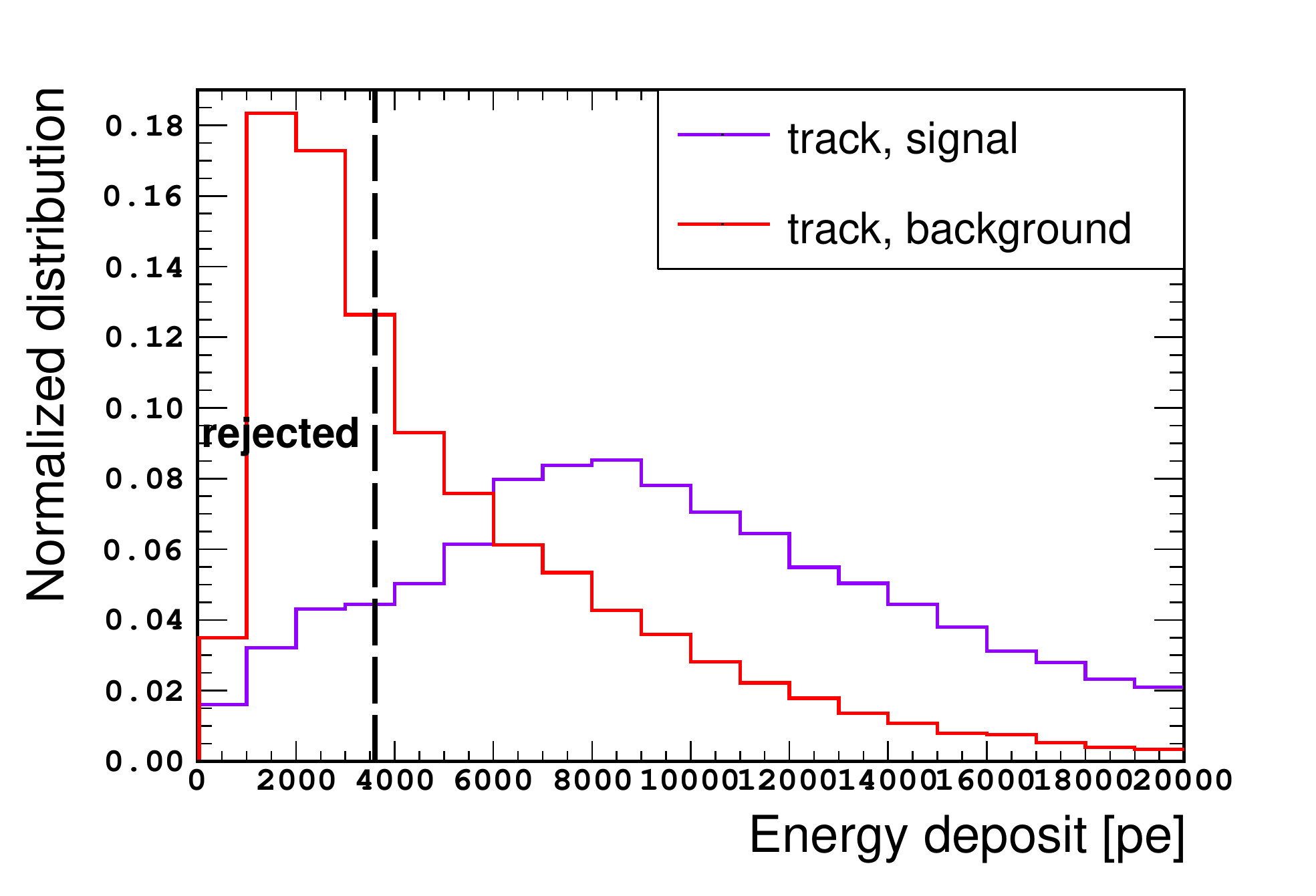}
\caption{The total energy deposit of the first object in time in the track case. Neutron-induced track deposits larger energy.}
\label{fig:edep_track}
\end{figure}


\section{Low-\texorpdfstring{$\nu$}{Lg} Analysis}
\label{sec:low_nu}
The low-$\nu$ method was proposed by Mishra~\cite{Mishra:1990ax} and has been used for neutrino and antineutrino charged-current flux, and cross-section measurements in the MINER$\nu$A experiment ~\cite{minerva_lownu}.
The peculiarity of the low-nu method is that the predicted cross section as a function of energy results to be flat for a certain cut on $\nu$\textcolor{black}{, which is the energy transfer to the nuclear system}. Assuming perfect knowledge of the detection efficiency and geometric acceptance, the shape of the low-$\nu$ sample energy spectrum is equal to the shape of \textcolor{black}{the} incoming neutrino flux. A good low-$\nu$ sample can provide correction and constraint on the neutrino flux.

On the other hand, the normalization in the low-$\nu$ region is rather unclear. The experimental handling is usually to take an external measurement of the high energy absolute cross section and scale the low-$\nu$ cross section normalization to it. This study is not taking this normalization into account.

The MINER$\nu$A experiment uses the calorimetric energy for the antineutrino low-$\nu$ channel study, which may \textcolor{black}{result in an underestimate of the neutrino energy~\cite{minerva_lownu}}. The 3DST is capable of obtaining information on each individual particle in the final state, including the neutron. Therefore a different energy transfer calculation method gives a hint of the usefulness of individual neutron kinematics detection.

\begin{figure}[h!]
\centering
\includegraphics[scale=0.45]{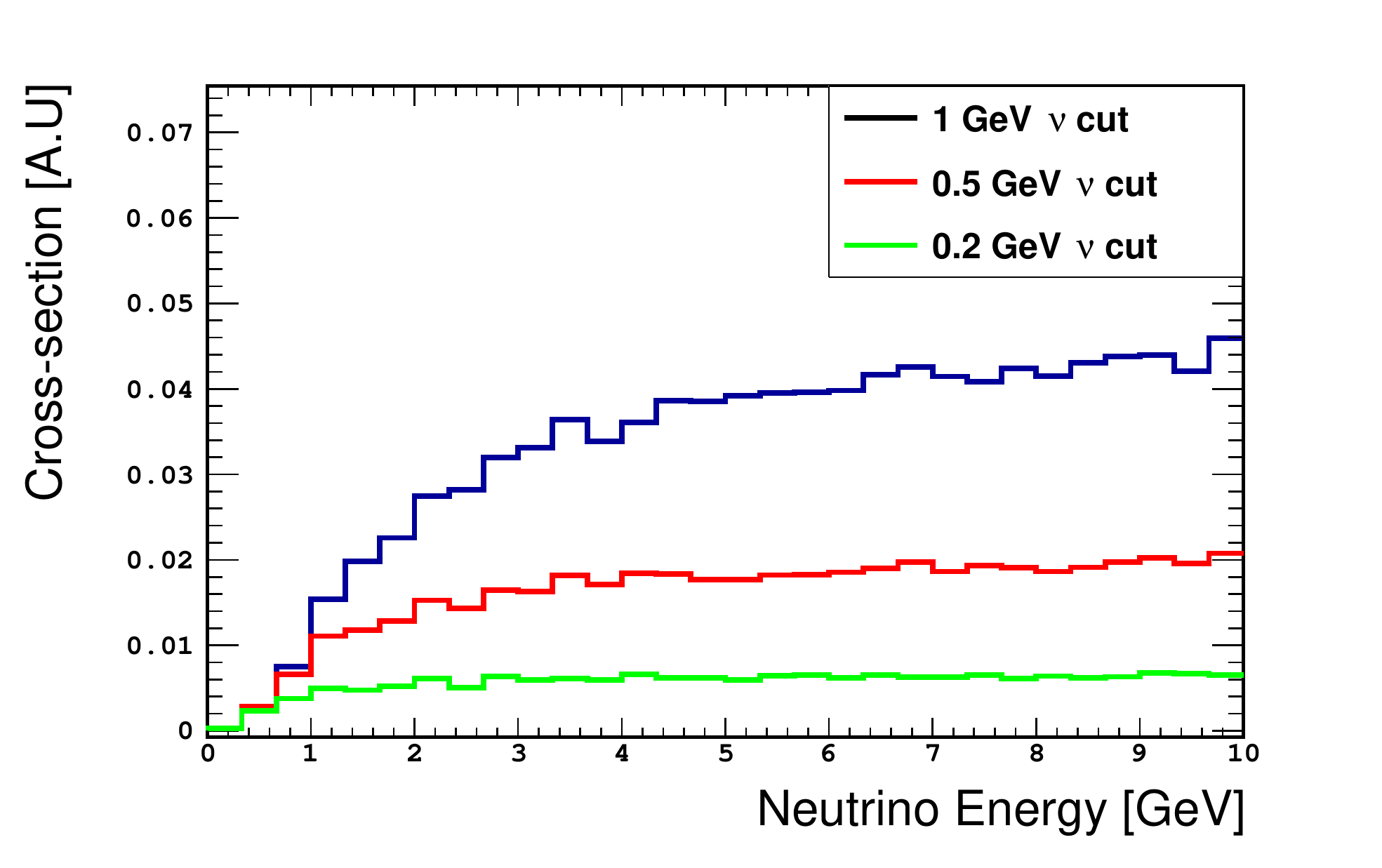}
\caption{Cross section shape as a function of $E_\nu$ with various $\nu$ selection cuts. The lower $\nu$ results in a flatter cross section.}
\label{fig:flat_cross_section}
\end{figure}

The CC inclusive cross section can be written as 
\begin{equation}
\begin{split}
\label{eq:cross_section}
\frac{d\sigma}{d\nu} = \frac{G^2_F M}{\pi} \int^1_0 \bigg( F_2 - \frac{\nu}{E_\nu} [F_2 \mp xF_3] \\
+ \frac{\nu}{2E^2_\nu} \bigg[ \frac{Mx(1-R_L)}{1+R_L} F_2 \bigg] \\
+ \frac{\nu^2}{2E^2_\nu} \bigg[ \frac{F_2}{1+R_L} \mp xF_3\bigg] \bigg) dx,
\end{split}
\end{equation}
where $E_\nu$ is the neutrino energy, $\nu$ is the energy transfer to the nuclear system and $G_F$ is Fermi constant \cite{Devan:2015uak}.
The cross section will be approximately constant as a function of $E_\nu$ if the $\nu$ is small enough compared to $E_\nu$ as shown in Fig.~\ref{fig:flat_cross_section}.
With proper efficiency and acceptance correction, the utilization of low-$\nu$ events results in a rather stringent antineutrino flux shape constraint since the measured neutrino spectrum shape directly reflects the flux shape.

The low-$\nu$ sample can be selected among the CC0$\pi$0p1n sample with the selection of reconstructed $\nu$ $<$ \SI{300}{MeV}.
The neutron's kinetic energy can be used as the reconstructed $\nu$ since, in the CC0$\pi$0p1n channel, the energy transfer to the nuclear system will go to the neutron, assuming that the binding energy of the nucleus is negligible.

\begin{figure}[h!]
\centering
\includegraphics[scale=0.45]{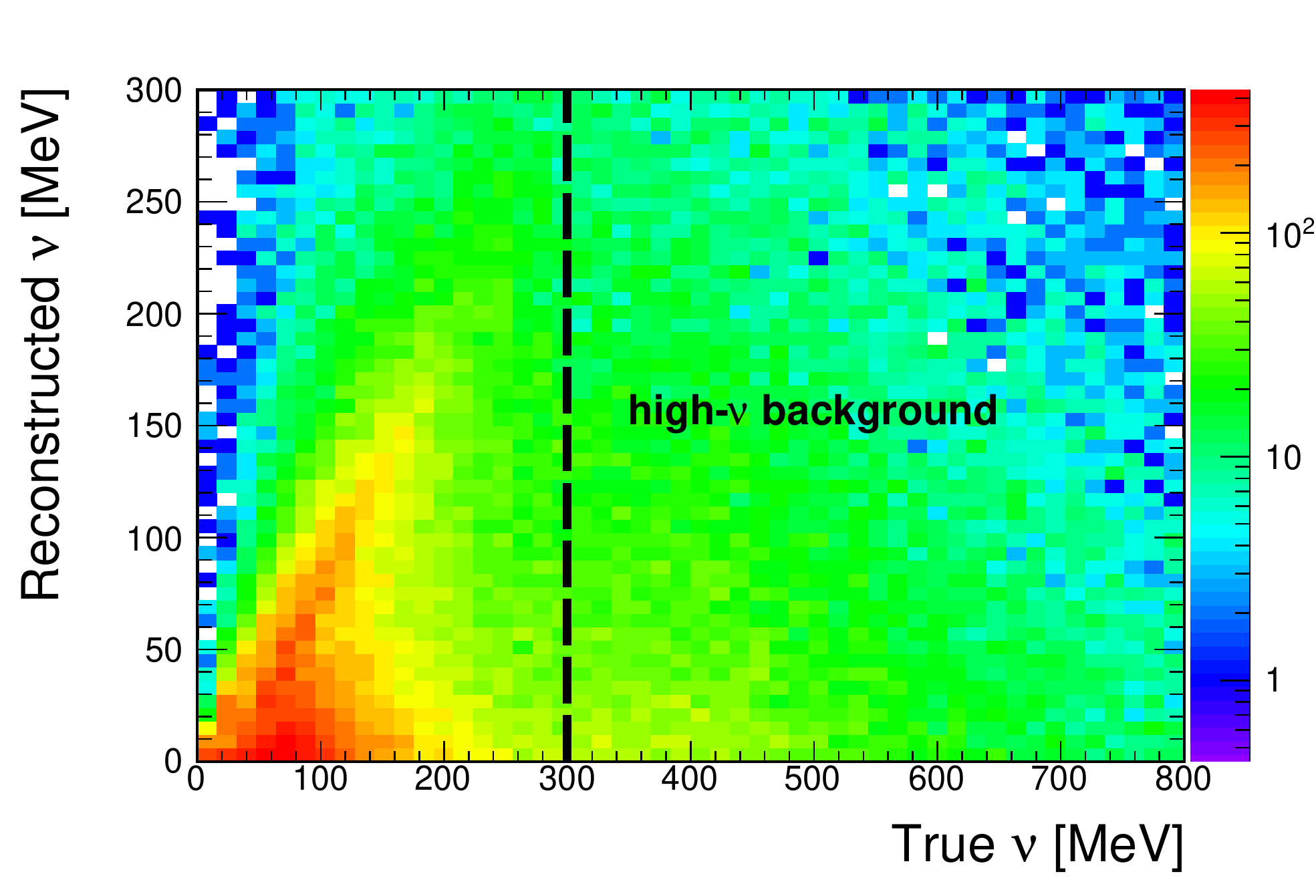}
\caption{Some events can have true $\nu$ larger than the low-$\nu$ cut (\SI{300}{MeV}); such events are defined as a high-$\nu$ background. The true $\nu$ is $E_\nu - E_\mu$ and the reconstructed $\nu$ is the measured kinetic energy of neutron by the ToF technique. The right region of the dashed line shows the high-$\nu$ background.}
\label{fig:high_nu}
\end{figure}
A ``high-$\nu$'' background shown in Fig.~\ref{fig:high_nu} should be rejected since it can make an undesired distortion of the desired flat cross section.
The main source of the high-$\nu$ background is events that have multiple neutrons in the final state.
Events with more than one neutron satisfy the low-$\nu$ cut even though they have a higher true $\nu$.
High-$\nu$ background can be reduced by \textcolor{black}{the} selection mentioned in Section~\ref{subsec:neutron_selection}.
Table.~\ref{table:2} shows the purity and efficiency of the low-$\nu$ sample.

\begin{table}[h!]
\centering
\begin{tabular}{ |p{4cm}||p{1.5cm}|p{1.5cm}|  }
 \hline
 \multicolumn{3}{|c|}{Purity and efficiency} \\
 \hline
 Cut& purity & efficiency \\
 \hline
ToF~(including threshold)  &   0.34  & 0.70\\
 energy deposit &   0.35  & 0.57\\
 branch number & 0.39 & 0.56\\
 max angle    & 0.64 & 0.30\\
 max distance & 0.66 & 0.26\\
 low-$\nu$ & 0.72  & 0.13\\
 \hline
\end{tabular}
\caption{purity and efficiency for each step of selection. The signal is CC0$\pi$0p1n low-$\nu$ events.}
\label{table:2}
\end{table}

\begin{figure}
\centering
\includegraphics[scale=0.45]{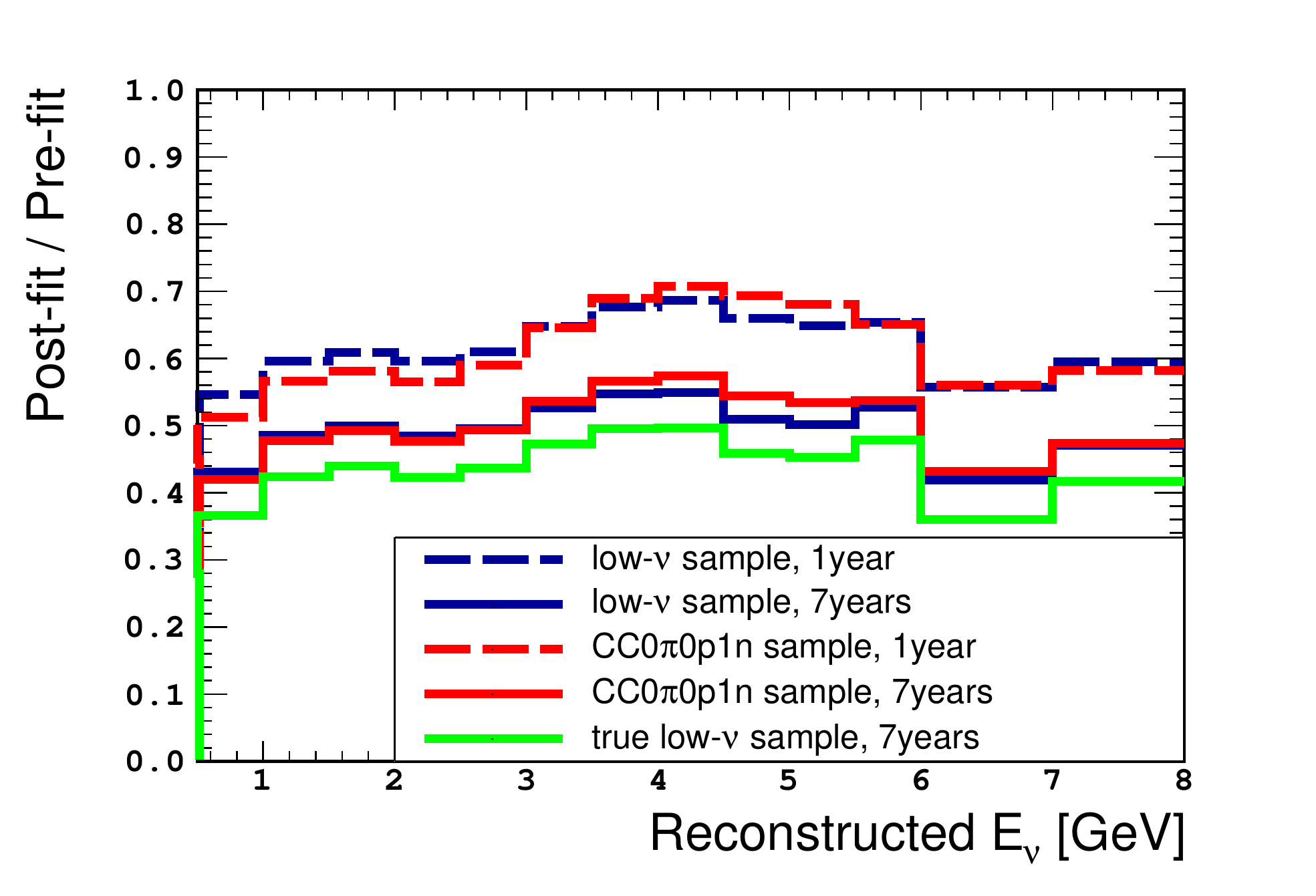}
\caption{Low-$\nu$ fitting result for flux uncertainty with various samples. It compares the quadrature sum of the flux uncertainty before and after the fit. With and without the low-$\nu$ selection with one year and seven years statistics, true CC0$\pi$0p1n low-$\nu$ sample with seven years statistics.}
\label{fig:lownu_result}
\end{figure}

The same $\chi^2$ fitting framework in Section~\ref{subsec:fitting} is used in this analysis.
The $\sigma_{f_B}$ can be constrained from 100\% to 85\% by sideband fitting. 
The high-$\nu$ backgrounds can be used as a sideband for the low-$\nu$ sample.
The flux uncertainty constraint with the low-$\nu$ method is presented by comparing the post-fit and pre-fit flux uncertainties as shown in Fig.~\ref{fig:lownu_result}.  

One important note is that according to Table.~\ref{table:2}, the low-$\nu$ cut reduces half of the statistics compared to \textcolor{black}{the} CC0$\pi$0p1n selection. This trade-off leads to an insignificant improvement of the flux constraint with the additional low-$\nu$ cut. Fig.~\ref{fig:lownu_result} shows such a trade-off effect. With the same running time, the overall constraints by a selected low-$\nu$ sample and \textcolor{black}{the} CC0$\pi$0p1n sample are similar.
At the low energy region, the selected CC0$\pi$0p1n sample without a flat cross section can also provide flux constraint due to relatively small cross-section uncertainty. 
Compared to the CC0$\pi$0p1n sample, additional constraint on the high-energy neutrino ($>$ \SI{3}{GeV}) due to the low-$\nu$ selection can be achieved. The overall flux constraint with the low-$\nu$ selection is shown in Fig.~\ref{fig:lownu_result}. 
However, due to the ineffectiveness of the low-$\nu$ method and loss of statistics, the constraint on the low energy region with the low-$\nu$ sample is less significant than the CC0$\pi$0p1n sample.

The low-$\nu$ method has a large model dependence since the low-$\nu$ cross section strongly depends on the modeling of the neutrino interaction.
There are possible models such as GiBUU, NEUT, NuWro, GENIE, etc., and GENIEv3 is used to model the interaction in this analysis.
As reported in~\cite{https://doi.org/10.48550/arxiv.2203.11821}, the shape of $\bar{\nu}_\mu-$C$_n$H$_n$ cross section spreads along the choice of the model, especially GENIEv2 and GENIEv3 10a configuration has the largest discrepancy at \SI{1}{GeV} $<$ E$_\nu < $ \SI{3}{GeV}.
Thus, the comparison of GENIEv2 and GENIEv3 is used to investigate the robustness of the low-$\nu$ method for the flux constraint. 
The model uncertainty is obtained by comparing the true CC0$\pi$0p E$_\nu$ cross section with the two models, and it's included as a systematic uncertainty for the diagonal term\textcolor{black}{s} of $M_{cov}$.
As shown in~\cite{https://doi.org/10.48550/arxiv.2203.11821}, the low-$\nu$ method is fragile to potentially large and not well-known systematic uncertainties due to the neutrino-nucleus interaction model. However, it is not the target of the present paper to evaluate such systematics, even if it is a crucial point that the community has to address to demonstrate if the low-$\nu$ method can be used reliably. Here we use the low-$\nu$ method only as an example to demonstrate the capability of the proposed detector design.

\newpage

\bibliographystyle{apsrev4-1}
\bibliography{reference}

\end{document}